\documentclass[aps,epsfig]{revtex4}
\usepackage{color}
\usepackage{graphicx}

\begin{document}
\def\be{\begin{equation}}
\def\ee{\end{equation}}     
\def\bfi{\begin{figure}}
\def\efi{\end{figure}}
\def\bea{\begin{eqnarray}}
\def\eea{\end{eqnarray}}

\title{Dynamic fluctuations in unfrustrated systems:
random walks, scalar fields and the Kosterlitz-Thouless phase}

\author{Federico Corberi$^{1,2}$ and Leticia F. Cugliandolo$^2$
\\
$^{1}$
{\small Dipartimento di Fisica ``E. R. Caianiello'', Universit\`a  di Salerno,}
\\ 
{\small via Ponte don Melillo, 84084 Fisciano (SA), Italy, and} 
\\
$^2$ 
{\small Universit\`e Pierre et Marie Curie, Paris VI,}
\\
{\small LPTHE UMR 7589, 4 Place Jussieu, 75252 Paris Cedex 05, France}
}

\today

\begin{abstract}

We study analytically the distribution of fluctuations of the quantities 
whose average yield the usual two-point correlation and linear 
response functions in three unfrustrated models: 
the random walk, the $d$ dimensional scalar field and the $2d$ XY model.
In particular we consider the time dependence 
of ratios between composite operators formed with these fluctuating quantities
which generalize the largely studied fluctuation-dissipation ratio, 
allowing us to  discuss the relevance of the effective temperature notion beyond linear order. The behavior of fluctuations in the 
aforementioned solvable cases is compared to numerical simulations of the $2d$ clock model with
$p=6,12$ states. 

\end{abstract}

\maketitle
\newpage

\tableofcontents

\newpage

\section{Introduction}

While in equilibrium systems the correct 
weighting of fluctuations has disclosed the basis of
modern statistical mechanics, 
uncovering the properties of non-equilibrium
fluctuations remains one of the most challenging and far reaching 
open questions of statistical physics.
However, in
spite of recent important developments \cite{Bertini}, explicit 
calculations, especially for interacting systems, are limited
to a few cases \cite{Touchette}. While for non-equilibrium stationary states 
a major advance has been the recognition of a general
symmetry of the probability distribution of certain observables, described
by the so-called fluctuation theorems \cite{Evans},
the behavior of fluctuations in slowly relaxing systems such as those
undergoing critical dynamics, phase-ordering kinetics, glassy 
evolution, forced relaxation and others, is much less understood.
Traditionally these systems
have been characterized in terms of the space and time dependence of 
two-point correlation functions and responses (see, {\it e.g.}, the review articles~\cite{Hoha,Bray,Cugliandolo02}). 
It has been only relatively recently that interest has moved to the study of fluctuations 
of these quantities and, in particular, their characterization 
{\it via} higher order correlation and response 
functions~\cite{ChamonCugliandolo,Cocuyo,noichi2,annibsollich,noivarianze,noispinglass}. 
These properties should, obviously, give us a more detailed description of the behavior 
of these highly non-trivial processes. A large activity around the analysis of fluctuations in 
glass forming liquids and other similar materials exists~\cite{book-fluct}.

In this paper we study analytically the distribution of fluctuations of some quantities 
the average of which yield the usual
two-point correlation function and linear response in three unfrustrated 
models: the random walk, the $d$ dimensional scalar field and the $2d$ XY model. In 
its structure this paper follows the presentation in~\cite{cuglia94}  where aging and 
the usual fluctuation-dissipation (FD) relation were studied in these same three models
with the purpose of highlighting the fact that neither disorder nor glassiness are needed to
obtain non-trivial dynamic features. Composite operators 
formed with the above-mentioned fluctuating quantities are related to
higher order correlation and response functions 
\cite{noichi2,noivarianze,noispinglass,Secumo}, and 
ratios between such composite operators can then be defined which
generalize the largely studied FD ratios.
The analysis of these ratios in the asymptotic time
limit inform us, in particular, on the relevance of the effective temperature notion~\cite{Cukup} (for reviews 
see~\cite{Crri,Coliza,Cugliandolo11}) beyond the {\it linear order}.
These analytical studies are complemented by
numerical simulations of the $2d$ clock model with $p=6$ and $p=12$ states, where a similar
behavior is found.

The paper is organized as follows. In Sec.~\ref{sec:fluc} we define the quantities 
that we will compute. Section~\ref{sec:random-walk} is devoted to the analysis of Brownian
motion in the over-damped limit. In Sec.~\ref{sec:scalar} we solve the Langevin dynamics 
of the scalar field model in $d$ dimensions. In Sec.~\ref{sec:clock} we study the evolution 
of the clock model and its limit, the XY model, in two dimensions where a 
Kosterlitz-Thouless (KT) phase exists in which the models are critical. Finally, in 
Sec.~\ref{sec:conclusions} we present our conclusions.

\section{Fluctuating quantities}
\label{sec:fluc}

We wish to characterize the statistics  of fluctuating quantities the average of which 
yield the two-time correlation and linear response. 
The choice  of these fluctuating quantities is not unique and 
their meaningful definition depends on the problem of interest, as we will discuss below. In the 
simpler cases (as the random walk of  Sec.~\ref{sec:random-walk} or the scalar field considered in 
Sec.~\ref{sec:scalar}) one is usually interested in the correlation and response 
of a field whose dynamics is ruled by a Langevin equation of the type
\be
\frac{\partial \phi (\vec x,t)}{\partial t}
=
-\Gamma 
\frac{\delta {\cal F}(\{\phi\})}{\delta \phi(\vec x,t)}+\xi (\vec x,t),
\label{langev}
\ee
${\cal F}$ being an effective Ginzburg-Landau free-energy 
and $\xi$ a thermal noise with the usual properties
$\langle \xi(\vec x,t) \rangle =0$ and 
$\langle \xi(\vec x,t) \xi(\vec x',t')\rangle =2 T \delta (\vec x-\vec x')\delta(t-t')$ 
(we set $k_B=1$ and we absorbed the coefficient $\Gamma $ with a redefinition of time).
This equation is complemented by the choice of an initial condition $\phi(\vec x,t=0)$.

In this stochastic dynamic equation there are two sources of fluctuations: the initial condition,
if taken from a probability distribution, and the random force mimicking the thermal noise. In the 
process of computing a response function, the applied perturbation may be an additional 
source of fluctuations, if it is taken to be random. In the analytical approaches contained in this
paper we will choose to work with 
fixed initial conditions and we will therefore freeze the first source of fluctuations mentioned above.
The effect of random initial conditions will be considered in Sec.~\ref{clockquench} 
for a discrete model, the clock model, which cannot be described in terms of a Langevin equation
and must be studied numerically.

Let us now introduce the two-time average quantities we will be interested in, the correlation and the linear 
response function, and their fluctuating parts. The two-time correlation function is defined as
\be
C \equiv C(\vec x,\vec x';t,t')
=\langle \phi(\vec x,t) \phi(\vec x',t')\rangle 
\; , 
\label{Cfirst}
\ee
where 
$\vec x$ and $\vec x'$ are two generic points in space and
$t$ and $t'$ two generic times. 
From this expression the definition of the fluctuating part of the correlation as
\be
\widehat C \equiv \widehat C(\vec x,\vec x';t,t')
=\phi(\vec x,t) \phi(\vec x',t') 
\; , 
\label{flucC-def}
\ee
is meaningful, since taking the average of the quantity
in Eq.~(\ref{flucC-def}) one immediately recognizes the (averaged) correlation
$C\equiv C(\vec x,\vec x';t,t')$. Although the definition (\ref{flucC-def}) is  
not unique, since other different fluctuating quantities may share the same average, 
for the correlation function this appears as the most {\it natural} choice.
Let us notice that not only the average of $\phi(\vec x,t) \phi(\vec x',t')$
yields $C(\vec x,\vec x';t,t')$, but composite operators (or moments) formed with this
fluctuating field provide higher order correlation functions. This observation,
which may seem quite trivial for the correlation function, is
not so obvious when dealing with the response function, for which the
definition of a fluctuating part is less straightforward.
The usual perturbation used to compute it is a field, say $h$, possibly drawn from a 
probability distribution, that couples linearly to the variable of interest, say $\phi $ itself, in such a way that 
${\cal F}\to {\cal F}-\int d^d x \ h(\vec x,t) \phi(\vec x,t)$. The 
perturbed Langevin equation then acquires an additional additive force. From here one  can easily 
prove~\cite{Cugliandolo02,Cugliandolo11,cuglia94}, 
for any ${\cal F}$, that the averaged linear response function is simply related to the correlation between $\phi$ and the 
noise $\xi$:
\begin{equation}
R(\vec x,\vec x'; t,t') = 
\left. \frac{\delta \langle \phi_h(\vec x,t)\rangle}{\delta h(\vec x',t')} \right|_{h=0}
= 
\frac{1}{2T} \langle \phi(\vec x,t) \xi(\vec x', t')\rangle
\; . 
\end{equation}
This relation gives us two fluctuating fields, 
$\frac{\delta \phi_h(\vec x,t)}{\delta h(\vec x',t')}|_{h=0}$
and $\phi(\vec x,t) \xi(\vec x', t')/(2T)$,
whose noise averages yield the linear response function, as {\it natural} candidates
to define the fluctuating part. The question then arises as to which is, between these two, the most interesting quantity to consider. 
In fact, although these objects have the same thermal average, they may have different fluctuation spectra and higher-order correlations.
In~\cite{noispinglass} we showed that the 
two-time fields appearing in the Martin-Siggia-Rose-Jenssen-deDominicis dynamic generating functional are 
naturally related to $\phi(\vec x,t) \xi(\vec x',t')/(2T)$. This suggests that 
the quantity $\phi(\vec x,t) \xi(\vec x', t')/(2T)$ may play a special physical role and is worth of being
studied. Indeed, special  and interesting properties are exhibited by this fluctuating quantity in the case
of aging spin-glasses \cite{noispinglass}. Similar considerations
have led to use this same fluctuating quantity in ferromagnetic systems in~\cite{noivarianze}.
Furthermore, it must be recalled that composite operators formed with the field
$\phi(\vec x,t)\xi (\vec x',t')/(2T)$ are directly related to higher order 
response functions \cite{noichi2,noivarianze,noispinglass,Secumo}, a property
analogous to the one discussed above for the fluctuating part of the correlation function.
This allows one to define 
ratios between composite operators of the fluctuating parts of response and 
correlation (see Sec.~\ref{sec:composite}), which are natural generalizations of the usual FD ratio and, by studying their behavior, 
to discuss the relevance of the notion of an effective temperature beyond linear order.
Finally, restricting to the quadratic models considered analytically in this paper,
the quantity $\frac{\delta \phi_h(\vec x,t)}{\delta h(\vec x',t')}|_{h=0}$ is deterministic,
while $\phi(\vec x,t) \xi(\vec x', t')/(2T)$ has a non-trivial fluctuating pattern.
Indeed, considering as an example a free-energy functional of the form
${\cal F}=\int d^dx \ (\nabla \phi)^2/2$ (but the argument is general for any quadratic form), one has that
the solution of Eq.~(\ref{langev}) is
\begin{equation}
\phi_h(\vec k,t) = \phi(\vec k,0) e^{-k^2t} + \int_0^t dt' \ e^{-k^2(t-t')} \ [\xi(\vec k,t') + h(\vec k,t') ] 
\end{equation}
from which one immediately concludes that
\begin{equation}
\frac{\delta \phi_h(\vec k,t)}{\delta h(\vec k',t')} = e^{-k^2(t-t)} \theta(t-t')
\end{equation}
is a deterministic function,  
independent of the initial condition, the applied field and the thermal noise realization. 
Instead, $\phi(\vec k,t) \xi(\vec k',t')/(2T)$ fluctuates and its higher order moments
are non-trivial, as we will show in the following. 

In conclusion, we 
define the fluctuating part of the linear response as 
\be
\widehat R\equiv\widehat R(\vec x,\vec x';t,t') =
 \phi(\vec x,t) \frac{\xi(\vec x',t')}{2T}
\; ,
\label{flucR-def}
\ee
which, together with Eq.~(\ref{flucC-def}), provide the definitions of fluctuations
of the two-time quantities that will be adopted throughout this paper.

 In the following we will also  consider the time-derivative of $\widehat C$:
\be
\partial_{t'}\widehat C=
\frac{\partial \widehat C}{\partial t'} (\vec x,\vec x';t,t')
\; ,
\ee  
and the integrated fluctuating linear response function 
\be
\widehat \mu=\widehat \mu(\vec x,\vec x';t,t',t'')=
\phi(\vec x,t)\int _{t'}^{t''}\frac{\xi(\vec x',z)}{2T}\,dz
\; .
\label{flucMu-def}
\ee
With the choice $t''=t$ the average of the latter quantity provides the
dynamic (sometimes denoted as zero field cooled) susceptibility  
\be
\chi (\vec x,\vec x';t,t')=
\langle \widehat\chi(\vec x, \vec x'; t,t')\rangle=
\int _{t'}^{t} ds \ \widehat R(\vec x, \vec x'; t,s)
\; . 
\ee
For the $2d$ XY model we will define the relevant fluctuating quantities 
in Sec.~\ref{sec:clock}. 

\subsection{Joint probability distribution}
\label{sec:joint}

In simple models with a quadratic Hamiltonian ${\cal H}$, the
probability distribution of the fluctuating quantities introduced above can be explicitly 
exhibited, and a full characterization of the fluctuation spectra can be given (for the $2d$ XY model an equivalent
delineation will be provided by studying the  moments of any order in
Sec.~\ref{sec:clock}). Let us start with the simplest case of a single time-dependent
function $\phi(t)$
(i.e. we consider a zero space dimensional problem; namely, there is no $\vec x$ dependence). 
It is convenient to introduce the quantities
\be
\Phi(t'',t')=\phi (t'')-\phi(t') 
\; ,
\label{flucGamma-def}
\ee
and
\be
\Xi (t'',t')=\int_{t'}^{t''}\xi (z)\,dz 
\; .
\label{flucGamma-def2}
\ee
In terms of these quantities the variation of the correlation 
$\Delta \widehat C=\widehat C(t,t'')-\widehat C(t,t')$ and 
the integrated response function $\widehat \mu $ read
\be
\Delta \widehat C(t,t'',t')=\phi (t)\,\Phi (t'',t') \; ,
\ee
and 
\be
\widehat \mu (t,t'',t')=\frac{\phi (t)\,\Xi (t'',t')}{2T} \; .
\ee
(With an abuse of language we omit to call ``fluctuating" the hat quantities.)
Notice also that the differential quantities $\partial _{t'}\widehat C$ and $\widehat R$
can be obtained from these forms as
$\partial _{t'}\widehat C=\lim _{\delta \to 0}\Delta \widehat C /\delta$ and
$\widehat R=\lim _{\delta \to 0}\widehat \mu /\delta$,
where $\delta =t''-t'$.
The joint probability distribution
${\cal P}(\phi,\Phi,\Xi)$ of having $\phi$ at time $t$,
$\Phi$ at times $t'$ and $t''$, and $\Xi$ also at times $t'$ and $t''$ is Gaussian and reads
\be
{\cal P}(\phi,\Phi,\Xi)=(2\pi)^{-\frac{3}{2}}\,\vert G\vert^{-\frac{1}{2}}
\exp \left [-\left (\begin{array}{ccc} \phi & \Phi & \Xi \end{array} \right )G^{-1}
\left (\begin{array}{c} \phi \\ \Phi \\ \Xi \end{array}\right) \right ]
\ee
where $G$ is the matrix of correlations:
\be
G=
\left ( \begin{array}{ccc} \langle \phi ^2 \rangle & 
\frac{1}{2} \Delta C&
\frac{1}{2}\mu\\
\frac{1}{2}\Delta C& 
\langle  \Phi ^2\rangle& 
\frac{1}{2}\langle \Phi  \Xi \rangle\\ 
\frac{1}{2} \mu &
\frac{1}{2}\langle  \Phi  \Xi \rangle& 
\langle \Xi ^2 \rangle \end{array} \right ) ,
\ee
$|G|$ its determinant, $\langle \phi^2\rangle=C(t,t)$ is the field correlator, 
$\langle \Phi^2\rangle=C(t'',t'') + C(t',t') - 2 C(t'',t')$ is the field ``displacement", 
$\Delta C = \langle \Delta \widehat C(t,t'',t')\rangle =C(t,t'')-C(t,t')$ and 
$\mu=\langle \widehat\mu \rangle 
=\int_{t'}^{t''} dz \ R(t,z)$ in the time-integrated linear response. 
For white noise $\langle \Xi^2 \rangle =2T (t''-t')$. 
Let us observe that, by choosing  $\phi, \Phi, \Xi$ as arguments of ${\cal P}$ one has
the advantage of having all finite entries in $G$ [this is not ensured if one uses, for instance, $\phi (t)$, $\phi (t')$
and $\xi (t')$]. 

In terms of ${\cal P}$, the joint correlation-response probability distribution
$P(\Delta \widehat C,\widehat \mu)$ reads 
\be
P(\Delta\widehat C\,,\,\widehat \mu)=
\int d\phi \, d\Phi \, d\,\Xi \,\, {\cal P}(\phi,\Phi,\Xi)\,
\delta [\phi \, \Phi -\epsilon \Delta \widehat C]\, \delta [\phi \,\Xi-2T\epsilon\widehat \mu]
\; .
\label{eqP}
\ee
Notice that we have introduced the parameter $\epsilon $ which allows us to
use a single form for the probability $P(\Delta \widehat C,\widehat \mu)$
of the integrated quantities, by simply setting $\epsilon=1$ in Eq.~(\ref{eqP}), and the probability
$P(\partial _{t'}\widehat C,\widehat R)$ of the differential ones
by letting $\epsilon=\delta$ and taking the limit $\delta \to 0$, namely  
\be
P(\partial _{t'}\widehat C,\widehat R)=
\left [ \lim_{\delta \to 0}\left . P(\Delta \widehat C,\widehat \mu)\right \vert _{\epsilon=\delta}\right ]_
{\stackrel{\widehat \mu=\widehat R}{\Delta \widehat C=\partial _{t'}\widehat C}}.
\label{preceipt}
\ee
Using the integral representation $\delta(x)=(2\pi)^{-1}\int _{-\infty}^\infty d\eta \,e^{-i\eta x}$ of
the Dirac function one arrives at
\begin{eqnarray}
P(\Delta \widehat C\,,\,\widehat \mu)
&=&
(2\pi)^{-\frac{3}{2}}\,\vert G\vert ^{-\frac{1}{2}} 
\int 
d\phi \, d\Phi \, d\Xi \;
\int _{-\infty+i\eta ^*}^{\infty +i\eta ^*} 
\frac{d\eta}{2\pi} \, 
\int _{-\infty+i\lambda ^*}^{\infty +i\lambda ^*} 
\frac{d\lambda}{2\pi} \, 
e^{-i(\eta \, \epsilon \Delta \widehat C+2\lambda T\epsilon \,\widehat \mu)}
\nonumber \\
&& 
\qquad\qquad\qquad\qquad\qquad\qquad \times 
\exp \left [-\left (\begin{array}{ccc} \phi & \Phi & \Xi \end{array} \right )B_{\eta \lambda}^{-1}
\left (\begin{array}{c} \phi \\ \Phi \\ \Xi \end{array}\right) \right ]
\label{pxxxnn}
\end{eqnarray}
with
\be
B_{\eta \lambda}^{-1}=G^{-1}+ \left ( \begin{array}{ccc} 0 & -i\frac{\eta}{2} & -i\frac{\lambda}{2}\\
 -i\frac{\eta}{2} & 0 & 0 \\
-i\frac{\lambda}{2} & 0 & 0 \end{array} \right )
.
\label{eqmatrixb}
\ee
Notice that, following standard techniques \cite{BerlinKac}, the integration paths in  Eq.~(\ref{pxxxnn}) have been 
deformed by the arbitrary shifts $\eta ^*$ and $\lambda ^*$. This does not change the value of
the integral since it merely amounts to
introducing an additive term $\eta ^*(\phi \Phi -\epsilon \Delta \widehat C)
+\lambda ^*(\phi \Xi -2T\epsilon \widehat \mu)$, which vanishes due to the 
$\delta$-constraints in Eq.~(\ref{eqP}), in the argument
of the exponential on the right-hand-side (rhs) of Eq.~(\ref{pxxxnn}).
In turn, a properly deformed integration path may render the integrations 
over $d\phi \,d\Phi \,d\Xi$ convergent in the cases in which (like in Sec.~\ref{secgaussprob}) the result over the
orginal route is divergent (when the integrations over $d\phi \,d\Phi \,d\Xi$ are taken 
before those over $d\eta \,d\lambda$).

Performing the Gaussian integrations in Eq.~(\ref{pxxxnn}) one obtains
\begin{eqnarray}
P(\Delta \widehat C\,,\,\widehat \mu)
&=&
\!\!
\vert G\vert ^{-\frac{1}{2}} \int _{-\infty+i\eta ^*}^{\infty+i\eta ^*}\frac{d\eta}{2\pi} \, 
\int _{-\infty+i\lambda ^*}^{\infty +i\lambda ^*}\frac{d\lambda}{2\pi} \,\, 
e^{-i(\eta \, \epsilon \Delta \widehat C+2\lambda T\epsilon \,\widehat \mu)}
\ \vert B_{\eta \lambda}\vert ^{\frac{1}{2}}\,\,
\nonumber \\
&=& \!\!
\int _{-\infty+i\eta ^*}^{\infty+i\eta ^*}\frac{d\eta}{2\pi} \, 
\int _{-\infty+i\lambda ^*}^{\infty+i\lambda ^*}\frac{d\lambda}{2\pi} \,\, e^{-i(\eta \, \epsilon \Delta \widehat C+2\lambda T\epsilon \,\widehat \mu)}
\left \{ 1-\frac{1}{2}i\eta \Delta C -\frac{1}{2}i\lambda (2T\mu) 
\right.
\nonumber\\
&&
\left.
\qquad\qquad
-\frac{1}{8}\eta \lambda [2T\mu \Delta C -2\langle \phi ^2\rangle \langle \Phi \Xi \rangle]
-\frac{1}{16}\eta ^2 [(\Delta C)^2-4\langle \phi^2 \rangle \langle\Phi ^2\rangle]
\right .\nonumber \\
&&\left .
\qquad\qquad
-\frac{1}{16}\lambda ^2 [(2T\mu)^2-4\langle \phi^2 \rangle \langle \Xi ^2\rangle]
\right \}^{-\frac{1}{2}}
.
\label{pesplicit}
\end{eqnarray}
Note that the three time-dependencies ($t > t'' > t'$) enter only 
{\it via} averages. We will explicitly compute $P$ in a specific zero-dimensional model,
the random walk, in Sec.~\ref{jointrw}.  
 
Let us now generalize what done insofar to a scalar field $\phi(\vec x,t)$ defined 
on a $d+1$ dimensional space.
Taking into account the space dependence the definitions (\ref{flucGamma-def}) and 
(\ref{flucGamma-def2})
are replaced by
\be
\Phi(\vec x',t'',t')=\phi (\vec x',t'')-\phi(\vec x',t')
\; ,
\ee
and
\be
\Xi(\vec x',t'',t') =\int_{t'}^{t''}\xi (\vec x',z)\,dz
\; .
\label{xi}
\ee
The fluctuating quantities we are interested in are
\begin{eqnarray}
\Delta \widehat C(\vec r=\vec x-\vec x',t,t'',t')&=&
\int d\vec x \ \phi (\vec x,t)\Phi(\vec x',t'',t')
 \nonumber \\
&=&\int \frac{d\vec k}{(2\pi)^d}\, \widetilde \phi(\vec k,t)\widetilde \Phi(-\vec k,t'',t')\,
e^{i\vec k (\vec x-\vec x')}\,\,,
\label{delC}
\end{eqnarray}
and 
\begin{eqnarray}
\widehat \mu(\vec r=\vec x-\vec x',t,t'',t')&=&
\frac{1}{2T}\int d\vec x \ \phi (\vec x,t)\Xi(\vec x',t'',t') 
\nonumber \\ 
&=&\frac{1}{2T}\int \frac{d\vec k}{(2\pi)^d}\, \widetilde \phi(\vec k,t)
\widetilde \Xi(-\vec k,t'',t')\,e^{i\vec k (\vec x-\vec x')}\,\,,
\label{mu}
\end{eqnarray} 
where $\widetilde \phi (\vec k,t)$ is the $\vec k$ component of the Fourier transform~\footnote{We use 
the following Fourier transform conventions 
$\phi(\vec x,t) = \int d^dk/(2\pi)^d \ e^{i\vec k \vec x} \,\widetilde \phi(\vec k,t)$
and 
$\widetilde \phi(\vec k,t) = \int d^dx \ e^{-i\vec k \vec x} \phi(\vec x,t)$.
We also use $\int d^dx e^{i\vec k (\vec x-\vec x')} = (2\pi)^d \delta(\vec x-\vec x')$.
} of the 
field $\phi (\vec x,t)$,
and similarly for $\widetilde \Phi$ and $\widetilde \Xi$. From here onwards $\vec r\equiv \vec x-\vec x'$.
When the $\vec k$ components are independent (i.e., for a quadratic Hamiltonian), 
the joint probability $P(\phi,\Phi,\Xi)$ is factorized as
\be
{\cal P}(\widetilde \phi,\widetilde \Phi, \widetilde \Xi)=\prod _k(2\pi)^{-\frac{3}{2}}\,\vert \widetilde G\vert^{-\frac{1}{2}}
\exp \left [-\left (\begin{array}{ccc} \widetilde \phi & \widetilde \Phi & \widetilde \Xi \end{array} \right )\widetilde G^{-1}
\left (\begin{array}{c} \widetilde \phi \\ \widetilde \Phi \\ \widetilde \Xi \end{array}\right) \right ]
\ee
where $\left (\begin{array}{ccc} \widetilde \phi & \widetilde \Phi & \widetilde \Xi \end{array} \right )$
and $\left (\begin{array}{c} \widetilde \phi \\ \widetilde \Phi \\ \widetilde \Xi \end{array}\right)$
are evaluated at the wavevectors $\vec k$ and $-\vec k$, respectively. 
$\widetilde G$ is the matrix of the $\vec k$-component correlations 
\be
\widetilde G=
\left ( \begin{array}{ccc} \widetilde C_{\phi \phi} & 
\frac{1}{2} \Delta \widetilde C&
\frac{1}{2}\widetilde \mu\\
\frac{1}{2}\Delta \widetilde C& 
\widetilde C_{\Phi \Phi}& 
\frac{1}{2}\widetilde C_{\Phi \Xi}\\ 
\frac{1}{2} \widetilde \mu &
\frac{1}{2}\widetilde C_{\Phi \Xi}& 
\widetilde C_{\Xi \Xi} \end{array} \right ) \,\,,
\ee
with
$\Delta \widetilde C=\widetilde C(\vec k,t,t'')-\widetilde C(\vec k,t,t')$, where 
\begin{equation}
(2\pi )^d\delta(\vec k+\vec k')\widetilde C(\vec k,t,t')=
\langle \phi (\vec k,t)\phi(\vec k',t')\rangle,
\label{ccdik}
\end{equation}
is the usual two-time structure factor. The other elements of the $\widetilde G$ matrix,
which as in the zero-dimensional case are all finite,
are the correlators 
$(2\pi )^d \delta (\vec k+\vec k') \widetilde \mu(\vec k,t,t'',t') =\langle \phi (\vec k,t)\Xi(\vec k',t'',t')\rangle$,
$(2\pi)^d \delta(\vec k + \vec k')\widetilde C_{\Phi \Xi}(\vec k,t,t'',t') =
\langle \widetilde \Phi (\vec k,t'',t')\widetilde \Xi(\vec k',t'',t')\rangle$, 
and similarly for $\widetilde C_{\phi \phi}$, $\widetilde C_{\Phi \Phi}$, and
$\widetilde C_{\Xi \Xi}$.

Starting from this,
the joint probability distribution $P(\Delta \widehat C,\widehat \mu)$
of the {\it real-space} quantities defined in Eqs.~(\ref{delC}) and (\ref{mu})
can be straightforwardly obtained proceeding along the same lines as for the zero-dimensional 
case. We first introduce the matrix 
$B_{\eta \lambda}$ which is related to $\widetilde G$ by
\be
B_{\eta \lambda}^{-1}(\vec k,\vec r=\vec x-\vec x')=\widetilde G^{-1}(\vec k)+ 
\left ( \begin{array}{ccc} 0 & -i\frac{\eta}{2} e^{i\vec k (\vec x-\vec x')} & 
-i\frac{\lambda}{2} e^{i\vec k (\vec x-\vec x')}\\
 -i\frac{\eta}{2} e^{i\vec k(\vec x-\vec x')} & 0 & 0 \\
-i\frac{\lambda}{2} e^{i\vec k (\vec x-\vec x')} & 0 & 0 \end{array} \right )\,\,,
\label{eqmatrixbgauss}
\ee  
and is the generalization of Eq.~(\ref{eqmatrixb}). We next obtain 
\begin{eqnarray}
P(\Delta \widehat C\,,\,\widehat \mu)
&=&
\int _{-\infty+i\eta^*}^{\infty+i\eta^*} \frac{d\eta}{2\pi} \, 
\int _{-\infty+i\lambda ^*}^{\infty +i\lambda ^*}
\frac{d\lambda}{2\pi} \,\, 
\exp\left[ -i(\eta \, \epsilon \Delta \widehat C+2\lambda T\epsilon \,\widehat \mu) + 
                 \frac{V}{2}\int \frac{d\vec k}{(2\pi)^d} \ln \left (
	        \frac{\vert B_{\eta \lambda}(\vec k,\vec r)\vert}{\vert G(\vec k)\vert}\right )
	        \right]
\nonumber \\
&=& 
\int _{-\infty+i\eta^*}^{\infty+i\eta ^*}\frac{d\eta}{2\pi} \, 
\int _{-\infty+i\lambda ^*}^{\infty +i\lambda^*}
\frac{d\lambda}{2\pi} \,\, 
\exp\left[ -i(\eta \, \epsilon \Delta \widehat C+2\lambda T\epsilon \,\widehat \mu) \right]
\nonumber\\
&&
\qquad\qquad
\times 
\exp\left\{
-\frac{V}{2}\int  \frac{d\vec k}{(2\pi)^d} \ln \left (
1-\frac{1}{2}i\eta e^{i\vec k \vec r}\Delta \widetilde C -\right . \right.
\nonumber \\
&&
\qquad\qquad\qquad
\frac{1}{2}i\lambda 
e^{i\vec k\vec r}(2T\widetilde \mu) 
-\frac{1}{8}\eta \lambda e^{2i\vec k\vec r}
[2T\widetilde \mu \Delta \widetilde C -2\widetilde C _{\phi \phi} 
\widetilde C_{\Phi \Xi}]- \nonumber \\
&&
\qquad\qquad\qquad 
\frac{1}{16}\eta ^2 e^{2i\vec k \vec r}
[(\Delta \widetilde C)^2-4\widetilde C_{\phi \phi} \widetilde C_{\Phi \Phi}]- 
\left . \left . \hspace{-.1cm}\frac{1}{16}\lambda ^2 e^{2i\vec k \vec r}
[(2T\widetilde \mu)^2-4\widetilde C_{\phi \phi} \widetilde C_{\Xi \Xi}]
\right )\right \},
\label{pgauss}
\end{eqnarray}
where $V$ is the volume of the system, which extends Eq.~(\ref{pesplicit}) to the finite dimensional
case. This form is totally general (for a quadratic Hamiltonian).  Note that the three time dependencies ($t > t'' > t'$) still enter only 
{\it via} averages, as in the zero-dimensional case, while the $\vec r$- (actually $r$-)
dependence is explicit.
We will discuss the properties of $P$ in a specific $d$-dimensional model, the
scalar field, in Sec.~\ref{secgaussprob}.

\subsection{Composite operators}
\label{sec:composite}

We now go on by using the generic notation and we build averages of composite fields of the form
\be
{\cal D}^{(n,m)}=
\langle 
\prod_{i=m+1}^n \partial _{t'_i} \widehat C_i \,
\prod _{j=1}^{m} \widehat R_j \; \rangle
\; ,
\label{momder}
\ee
where we have used the shorthand
\be
\widehat C_i=\widehat C(\vec x_i,\vec x'_i;t_i,t'_i)
\ee
and similarly for $R_i$, where $\vec x_i$ and $\vec x'_i$ ($t_i$ and $t'_i$) are two generic
space positions (times), with $t_i\ge t'_i$. 

It is also interesting to use the quantity in which the fluctuating time-variation of the 
correlation and linear response have been  time-integrated:
\be
{\cal C}^{(n,m)}=\langle \prod_{i=m+1}^n \widehat C_i \,
\prod_{j=1}^{m} \widehat \chi_j \; \rangle
\; .
\label{momC}
\ee
Moments are obtained from these quantities by subtracting
a suitable disconnected part, as will be discussed in Sec.~\ref{moments}. 
The averaged correlation, linear response and the integrated linear response are simply 
${\cal C}^{(1,0)}$, ${\cal D}^{(1,1)}$ and ${\cal C}^{(1,1)}$, respectively. 
Analogously, for $n>1$ the quantities ${\cal C}^{(n,0)}$ (or ${\cal D}^{(n,0)}$) are
higher order  correlation functions (or their time derivatives) and,
similarly, as discussed in \cite{noichi2,noivarianze,noispinglass,Secumo},  
${\cal C}^{(n,n)}$ and ${\cal D}^{(n,n)}$ are related to higher order 
response functions (time integrated or impulsive, respectively) 
\footnote{A caveat applies to the case $(\vec x_i,t_i)=(\vec x_i',t_i')$. See \cite{noichi2,noivarianze} for a discussion.}.

Notice that the quantities in Eqs.~(\ref{momder}) and (\ref{momC}) cannot be obtained from the
joint probability distribution computed in Sec.~\ref{sec:joint}: indeed, in general,
they involve correlations and responses evaluated at different space-time variables 
$\vec x_i,\vec x'_i;t_i,t'_i$ for any $i$, whereas both $\Delta \widehat C$ and $\widehat \mu$ 
in Eqs.~(\ref{delC}) and (\ref{mu}) are considered with the same space-time arguments 
(namely, $\vec x,\vec x'$ and $t,t',t''$).

Let us now define the generalized FD ratio as
\be
X^{(n;m,m')} \equiv T^{m-m'}\,\frac{{\cal D}^{(n,m)}}{{\cal D}^{(n,m')}}
\;.
\label{xmom-def}
\ee       
Letting $n=1$, $m=1$, $m'=0$ one recovers the usual FD ratio $X$.   
These quantities do not put higher order FDTs to the test directly as these are complicated
functions, involving several terms~\cite{Secumo,noichi2}, but they do evaluate whether 
all these terms scale in the same way thus allowing for the existence of an effective 
temperature taking finite values over non-trivial time regimes. 

\section{Brownian motion}
\label{sec:random-walk}

The overdamped Langevin dynamics of a Brownian particle  
is ruled by Eq.~(\ref{langev}) with a single function $\phi $ 
(i.e. there is no dependence on $\vec x$) and ${\cal H}\equiv 0$.  
In this case $\phi $ should be interpreted as the position of a 
Brownian particle on a line. The extension to the case of a vector $\vec \phi$ with
$d$ components, describing diffusion in $d$ dimensions, is trivial 
and the basic results remain unaltered.

\subsection{Joint probability distribution}
\label{jointrw}

In this simple problem one trivially has $\Phi\equiv \Xi$. The joint probability distribution 
(\ref{pesplicit}) of $\Delta \widehat C(t,t''t,')$ and $\widehat \mu(t, t'',t')$ then reduces to
\begin{eqnarray}
&& 
P(\Delta \widehat C\,,\,\widehat \mu) =
 \int_{-\infty}^\infty \frac{d\eta}{2\pi} \, \int_{-\infty}^\infty\frac{d\lambda}{2\pi} \,
 e^{-i(\eta \, \epsilon \Delta \widehat C+2\lambda T\epsilon \,\widehat \mu)}
 \,\nonumber \\
&&  
\qquad\qquad\qquad\qquad \times \frac{1}
{ \sqrt{1-\frac{1}{2}i(\eta +\lambda)  \Delta C
-\frac{1}{16}(\eta +\lambda )^2[(\Delta C)^2-4\langle \phi^2 \rangle \langle\Phi ^2\rangle]
}}
\label{pesplicitrw}
\end{eqnarray} 
with $\Delta C=2T(t''-t')$, $\langle \phi^2\rangle =2Tt$ and 
$\langle \Phi^2\rangle = \langle \Xi^2\rangle =  2T(t''-t')$, where we assumed that $t>t'' > t'$
and set $\eta ^* = \lambda ^*=0$ since every integral is convergent.

This form is symmetric under the exchange $\Delta \widehat C \leftrightarrow 2T\widehat \mu$,
thus indicating that these two quantities are equally distributed. Indeed, since $\Phi=\Xi$
one has $\Delta\widehat C = 2T \widehat \mu$ and, 
therefore, the fluctuations of the composite fields whose averages are the correlation and
linear response are just {\it identical} in this case. As a consequence, all the composite 
operators of Eq.~(\ref{momder}) 
with the same value of $n$ [with $(t_i,t'_i)=(t,t')$ $\forall i$]
scale in the same way (that is to say, they have the same dependence on the times
$t,t'$). Moreover, introducing $\omega =\eta + \lambda$ one can explicitly integrate 
expression (\ref{pesplicitrw}) and find
\begin{equation}
P(\Delta \widehat C\,,\,\widehat \mu) =
\delta \left(\epsilon(2T\widehat \mu-\Delta \widehat C)\right)
\int_{-\infty}^\infty \frac{d\omega }{2\pi} \frac {e^{-i\omega \epsilon \Delta \widehat C}}
{\sqrt{1-\frac{1}{2}i\omega \Delta C
-\frac{1}{16}\omega^2[(\Delta C)^2-4\langle \phi^2 \rangle \langle\Phi ^2\rangle]}}
\; . 
\end{equation}
In the above integral, the integrand has
two branch points at 
$\omega=\omega _{\pm}$, with
\begin{equation}
\omega_\pm=4\,\frac{\pm 2 \sqrt{\langle \phi^2 \rangle \langle\Phi ^2\rangle}-\Delta C}
{(\Delta C)^2-4\langle \phi^2 \rangle \langle\Phi ^2}
\; . 
\end{equation}
Performing the integral we obtain
\begin{equation}
P(\Delta \widehat C\,,\,\widehat \mu) =
\frac{4e^{\displaystyle{\frac{4\Delta C \epsilon \Delta \widehat C}{4\langle \phi^2 \rangle 
\langle\Phi ^2\rangle-(\Delta C)^2}}}}{\pi \sqrt{4\langle \phi^2 
\rangle \langle\Phi ^2\rangle-(\Delta C)^2}} \,
K_0\left [8\epsilon 
\frac{\sqrt{\langle \phi^2 \rangle \langle\Phi ^2\rangle}}
{4\langle \phi^2 \rangle \langle\Phi ^2\rangle-(\Delta C)^2}\,\vert \Delta \widehat C \vert \right ]
\delta \left(\epsilon(2T\widehat \mu-\Delta \widehat C)\right),
\end{equation}
where $K_0 (z)$ is the modified Bessel function which can be expressed as 
$K_0 (z)=\int _1 ^\infty e^{zx}(x^2-1)^{-1/2}$. This result is very close to the one presented 
in~\cite{Chcuyo} for the probability distribution function (pdf) of the two-time composite field 
$\phi_\alpha(\vec x,t) \phi_\alpha(\vec x,t')$
in the $O(N)$ ferromagnetic model in the large $N$ limit. In both cases, the time dependencies 
enter only through the correlation functions, $\langle \phi^2\rangle(t) = C(t,t)$,  
$C(t'',t'')$, $C(t',t')$ and $C(t'',t')$.

\subsection{Composite operators}

The properties of  
${\cal D}^{(n,m)}$ can be computed explicitly as
\begin{eqnarray}
{\cal D}^{(n;m)} &=& \langle \ \prod_{i=m+1}^{n}  \phi(t_i) \dot \phi(t_i') \prod_{j=1}^{m} \phi(t_j) \xi(t_j')/(2T) \ \rangle
\nonumber\\
&=&
\langle \ \prod_{i=m+1}^{n}  \phi(t_i) \xi(t_i') \prod_{j=1}^{m} \phi(t_j) \xi(t_j')/(2T) \ \rangle
\nonumber\\
&=&
\langle \ \prod_{k=1}^{n}  \phi(t_k) \xi(t_k') \ \rangle \; (2T)^{-m}
\; . 
\end{eqnarray}
The remaining average can be expanded in products of two-point correlation and linear response 
functions by using Wick's theorem applied to the Gaussian variables $\phi$ and $\xi$. In so doing one finds
the explicit $2n$-time dependence of ${\cal D}^{(n,m)}$. If one is interested in the behavior of the 
generalized FD ratio $X^{(n;m,m')}$ this calculation is not necessary since 
the non-trivial factors in the numerator and denominator cancel out and one simply finds
a constant, 
\begin{eqnarray}
X^{(n;m,m')} =  \left(\frac{1}{2} \right)^{m-m'}
\; , 
\label{xrw}
\end{eqnarray}
independently of $n$. For $m=1$ and $m'=0$ one recovers $X=1/2$, the usual 
FD ratio~\cite{cuglia94,Pottier03} of the 
random walk. Equation~(\ref{xrw}) generalizes the result of Sec.~\ref{jointrw}, showing that moments
with different $m$ but the same $n$ scale in the same way for {\it any choice} of the time variables. 

\section{The free scalar field}
\label{sec:scalar}

The Langevin relaxation of the scalar field in $d$ spatial dimensions is given by
\begin{equation}
\frac{\partial \phi(\vec x,t)}{\partial t} = - \frac{\delta F[\phi]}{\delta \phi(\vec x,t)} + \xi(\vec x,t) 
\; . 
\label{eq:Langevin-scalar}
\end{equation}
In the free-field case the Ginzburg-Landau functional is simply 
\begin{equation}
F[\phi] = \frac{1}{2} \int d\vec x \ [\nabla \phi(\vec x,t)]^2 
\label{eq:free-energy-scalar}
\; .
\end{equation}
The expectations of the thermal noise are the usual
ones reported below Eq.~(\ref{langev}).
Equations~(\ref{eq:Langevin-scalar}) and (\ref{eq:free-energy-scalar}) also constitute the 
Edwards-Wilkinson model for the motion of an interface (with no overhangs) in $d$ transverse 
dimensions. In the context of interfaces,  the fluctuations of a two-time quantity the average of which is the 
roughness were studied in~\cite{Bustingorry0,Bustingorry,Iguain,Pleimling}. 

The Fourier transformed noise statistics are such that 
$\langle \xi(\vec k,t) \rangle =0$ and 
$\langle \xi (\vec k,t)\xi (\vec k',t')\rangle $$ =
2T (2\pi)^d
\delta(\vec k+\vec k')\delta(\vec t-\vec t')$. 
Starting from $\phi(\vec x,0)=0$, without loss of
generality, one has
\be
\phi(\vec k,t) = 
\int_0^t ds \; e^{-k^2 (t-s)} \, \xi(\vec k,s)
\label{eq:solution-scalar}
\ee
and $\phi(\vec k,t)$ as well as $\phi(\vec x,t)$ inherit Gaussian statistics from $\xi$.
From Eq.~(\ref{eq:solution-scalar}) for the Fourier space correlator (\ref{ccdik}) one obtains
\be
\widetilde C(\vec k;t,t') =
\frac{T}{k^2}\left [e^{-k^2(t-t')}-e^{-k^2(t+t')}\right ] 
\; .
\label{ckgauss}
\ee

Introducing a short-distance cut-off $a^2=1/\Lambda ^2$ mimicking a lattice spacing,
so that 
$\int d\vec k \to \int d\vec k \ \exp(-\frac{k^2}{\Lambda ^2})$, 
the real space correlation function reads
\begin{equation}
C(r;t,t') 
\equiv 
\langle \, \phi(\vec x, t) 
\phi(\vec x', t') \, \rangle 
=
\int 
\frac{d\vec k}{(2\pi)^{d}}
\
\widetilde C(\vec k;t,t') 
\
e^{-\frac{k^2}{\Lambda ^2}} 
\
e^{-i\vec k \vec r}
\; , 
\end{equation}
with $r=|\vec x-\vec x'|$, and one finds
\begin{eqnarray}
C(r;t,t') =
\frac{Tr^{2-d}}{4\pi^{d/2}} \, 
\Gamma\left[\frac{d}{2}-1,\frac{\Lambda ^2r^2}{4[1+\Lambda^2 (t+t')]}, 
\frac{\Lambda^2 r^2}{4[1+\Lambda^2 (t-t')]} \right]
\; ,
\label{eq:gamma-scalar}
\end{eqnarray}
where 
\begin{equation}
\Gamma[n,a,b] \equiv \int_a^b dz \, z^{n-1} e^{-z}   
\end{equation}
is the generalized incomplete Gamma function. 
Analogously, the linear response is
\begin{eqnarray}
R(r;t,t')\equiv 
\left.
\frac{\delta \langle \phi(\vec x,t) \rangle_{h}}
{\delta h(\vec x', t')}  
\right|_{h=0} &=&
\langle \phi(\vec x,t) \xi(\vec x', t')\rangle/(2T)
\nonumber\\
&=&
\frac{\Lambda^d}{(4\pi)^{d/2}} \; 
\frac{e^{-\left[\displaystyle{\frac{\Lambda ^2 r^2}{4[1+\Lambda^2 (t-t')]}}\right]}
}
{\left [1+\Lambda^2 (t-t')\right ]^{d/2}} 
\, \theta(t-t')
\; .
\label{eq:bare-R0-scalar}
\end{eqnarray}
The relevant long-times limit is such that $\Lambda ^2(t-t')\gg 1$.
In this limit the partial derivative of Eq.~(\ref{eq:gamma-scalar}) with respect to $t'$ becomes 
\be
\partial _{t'}C(r;t,t')\simeq 
\frac{T\Lambda ^d}{2^{d}\pi^{d/2}} \, t'^{-d/2}\left [ \left ( y-1 \right )^{-d/2} e^{-\zeta}+
\left ( y+1 \right )^{-d/2} e^{-\frac{y-1}{y+1}\,\zeta} \right ]
\; ,
\label{eq:bare-dC-scalar}
\ee
(and similarly for $\partial_t C$) while from Eq.~(\ref{eq:bare-R0-scalar}) one has
\be
R(r;t,t')=
\frac{\Lambda ^d}{2^{d}\pi^{d/2}} \, t'^{-d/2} \left (y-1 \right )^{-d/2}  e^{-\zeta}
\, \theta(t-t') 
\; ,
\label{eq:bare-R01-scalar}
\ee
where $\zeta = r^2/[4(t-t')]$ and $y=t/t'$.
Equations~(\ref{eq:bare-dC-scalar}) and (\ref{eq:bare-R01-scalar}) mean that 
for $r^2\ll 4(t-t')$,
$R $ and $\partial _{t'}C$ scale in the same way, namely 
$R \simeq t'^{-d/2}f_{R} (t/t')$ and 
$\partial _{t'}C \simeq t'^{-d/2}f_{\partial C}(t/t')$, with 
$f_{R}(y)=\Lambda ^d[4\pi (y-1)]^{-d/2}$ and  
$f_{\partial C}(y)=T\Lambda ^d(4\pi)^{-d/2} [(y-1)^{-d/2}+(y+1)^{-d/2}]$, respectively.
In this regime, the FD ratio~\cite{leticia}
\be
\lim_{r^2 \ll (t-t')}
X(r;t,t')=
\lim_{r^2 \ll (t-t')}
\frac{R(r;t,t')}{\partial _{t'}C (r;t,t')}=
\left [1+\left (\frac{y-1}{y+1}\right )^{d/2}\right ]^{-1}
\label{xtheta}
\ee
is independent of $r$ and converges, for $y\to\infty$, to the limiting 
value~\cite{godluck}
\be
X_\infty \equiv \lim_{t/t'\to\infty } \ \lim_{r^2 \ll 2 \Lambda^2 (t-t')} X(r;t,t') =\frac{1}{2}.
\label{xthetalim}
\ee
Notice that this asymptotic value does not depend upon the distance for any choice
of $r$ [not only for  $r^2\ll (t-t')$]
since, for $y \gg 1$, $R$ and $\partial _{t'}C$
are proportional in any case:
\be
\lim_{t/t'\to\infty } X(r;t,t') =\frac{1}{2}
\; .
\label{xthetalim-gen}
\ee
This result is the same as the one found for the 
random walk problem, see Sec.~\ref{sec:random-walk}. 

\subsection{Joint probability distribution} \label{secgaussprob}

In the large volume limit the joint probability (\ref{pgauss}) can be computed by using saddle point
techniques. Changing the integration variables to $z_C=i\eta$, $z_\mu=i\lambda$,
and letting $z^*_C=i\eta^*, z^*_\mu=i\lambda ^*$, 
Eq.~(\ref{pgauss}) can be cast as 
\be
P(\Delta \widehat C\,,\,\widehat \mu)
=
\int _{-i\infty+z^*_C} ^{i\infty+z^*_C} 
\frac{dz_C}{2\pi} \, 
\int _{-i\infty+z^*_\mu} ^{i\infty+z^*_\mu} 
\frac{dz_\mu}{2\pi} \,\, 
e^{Vh(\widehat d_C,\widehat d_\mu;\vec x, \vec x' ; t,t',t'',z_C,z_\mu)}
\label{befsad}
\ee
where $\widehat d_C =\frac{\epsilon \Delta \widehat C}{V}$, 
$\widehat d_\mu =\frac{2T\epsilon \widehat \mu}{V}$ are the correlation and response
densities, and
\begin{equation}
h(\widehat d_C,\widehat d_\mu;\vec x, \vec x' ; t,t',t'',z_c,z_\mu)=-z_C \widehat d_C -z_\mu \widehat d_\mu 
+G(\vec x,\vec x',t,t',t'',z_C,z_\mu)
\end{equation}
with
\be
G(\vec x,\vec x',t,t',t'',z_C,z_\mu)=
-\frac{1}{2}\int  \frac{d\vec k}{(2\pi)^d} \ln H(\vec k,\vec x,\vec x'; t,t',t'',z_C,z_\mu)
\label{eqforg}
\ee
and
\begin{eqnarray}
&& H(\vec k,\vec x,\vec x'; t,t',t'',z_C,z_\mu)
=
1-\frac{1}{2}z_C e^{i\vec k|\vec x-\vec x'|}\Delta \widetilde C -
\frac{1}{2}z_\mu 
e^{i\vec k|\vec x-\vec x'|}(2T\widetilde \mu) 
\nonumber\\
&&
\qquad\qquad\qquad
+\frac{1}{8}z_C z_\mu e^{2i\vec k|\vec x-\vec x'|}
[2T\widetilde \mu \Delta \widetilde C -2\widetilde C _{\phi \phi} 
\widetilde C_{\Phi \Xi}] 
+\frac{1}{16}z_C ^2 e^{2i\vec k|\vec x-\vec x'|}
[(\Delta \widetilde C)^2-4\widetilde C_{\phi \phi} \widetilde C_{\Phi \Phi}]
\nonumber\\
&& 
\qquad\qquad\qquad
+ 
\frac{1}{16}z_\mu ^2 e^{2i\vec k|\vec x-\vec x'|}
[(2T\widetilde \mu)^2-4\widetilde C_{\phi \phi} \widetilde C_{\Xi \Xi}]
\; .
\label{eqforh}
\end{eqnarray}
 
In the large $V$ limit, to any choice of the fluctuating correlation and response 
$\widehat d_C$ and $\widehat d_\mu$ there corresponds a couple of real quantities 
$z_C=z^*_C$ and $z_\mu=z^*_\mu$ whose contribution
dominates the whole double integral in Eq.~(\ref{befsad}). 
$z^*_C(\vec x,\vec x',t,t',t'')$ and $z^*_\mu(\vec x,\vec x',t,t',t'')$ are solutions to the coupled system
of equation:
\begin{eqnarray}
\left \{
\begin{array}{rcl}
\displaystyle 
\widehat d_C 
&=&
\left . 
\displaystyle
\frac{\partial G}{\partial z_C}\right \vert _{z^*_C,z^*_\mu} 
\\
&=&
\displaystyle
\frac{1}{2}\int \frac{d\vec k}{(2\pi)^d} \frac{\frac{1}{2}e^{i\vec k|\vec x-\vec x'|}\Delta \widetilde C
-\frac{1}{8}e^{2i\vec k|\vec x-\vec x'|}
[2T\widetilde \mu \Delta \widetilde C -2\widetilde C _{\phi \phi} 
\widetilde C_{\Phi \Xi}]z^*_\mu 
-\frac{1}{8}e^{2i\vec k|\vec x-\vec x'|}
[(\Delta \widetilde C)^2-4\widetilde C_{\phi \phi} \widetilde C_{\Phi \Phi}]z^*_C}
{H(\vec k,\vec x,\vec x',t,t',t'',z^*_C,z^*_\mu)}
\; , 
\\
\displaystyle
\widehat d_\mu
&=&
\left . 
\displaystyle
\frac{\partial G}{\partial z_\mu}\right \vert _{z^*_C,z^*_\mu}
\\
&=&
\displaystyle
\frac{1}{2}\int \frac{d\vec k}{(2\pi)^d} \frac{\frac{1}{2}e^{i\vec k|\vec x-\vec x'|}2T \widetilde \mu
-\frac{1}{8}e^{2i\vec k|\vec x-\vec x'|}
[2T\widetilde \mu \Delta \widetilde C -2\widetilde C _{\phi \phi} 
\widetilde C_{\Phi \Xi}]z^*_C
-\frac{1}{8}e^{2i\vec k|\vec x-\vec x'|}
[(2T \widetilde \mu)^2-4\widetilde C_{\phi \phi} \widetilde C_{\Xi \Xi}]z^*_\mu}
{H(\vec k,\vec x,\vec x',t,t',t'',z^*_C,z^*_\mu)}
\; .
\end{array}
\right .
\label{saddle}
\end{eqnarray}
Once these equations are solved to find $z^*_C$ and $z^*_\mu$ the joint probability distribution
for large $V$ can be written as
\be
P(\Delta \widehat C\,,\,\widehat \mu)=
\frac{1}{(2\pi)^2} \,\, 
e^{Vh(\widehat d_C,\widehat d_\mu; \vec x, \vec x' ; t,t',t'',z^*_C,z^*_\mu)}
\; .
\label{aftsad}
\ee 
In order for Eq.~(\ref{eqforg}) to be defined and the whole procedure to be meaningful
the saddle point solutions must satisfy the constraint 
$H(\vec k,\vec x,\vec x'; t,t',t'',z^*_C,z^*_\mu)>0$. For any choice of 
$\vec k,\vec x,\vec x'; t,t',t''$, this defines the interior of the
ellipses $H=0$ in the $z^*_C,z^*_\mu$ plane. Since momenta are integrated
over, the constraint must be obeyed for all the values of $\vec k$.
In order to see which is the momentum which provides the most stringent condition
we must know the expression of all the correlators entering $H$ in Eq.~(\ref{eqforh}),
that we derive below.
Using Eq.~(\ref{eq:solution-scalar}) and the properties of the thermal noise,
with the definitions of the momentum-space correlators given below 
Eq.~(\ref{ccdik}), one readily finds
\begin{eqnarray} 
\widetilde \mu(\vec k; t,t'',t') 
&=&
\frac{1}{k^2}\left [e^{-k^2(t-t'')}-e^{-k^2(t-t')}\right ],
\label{tildemu}
\\
\widetilde C_{\Xi \Xi}(\vec k,t',t'')
\label{tildechi}
&=&
2T(t''-t')
\; ,
\\
\widetilde C_{\Phi \Xi}(\vec k,t',t'')
\label{tildec}
&=&
\frac{2T}{k^2}\left [ 1-e^{-k^2(t''-t')}\right ].
\end{eqnarray}
The other Fourier-space correlators are obtained using Eq.~(\ref{ckgauss}) and they read
\begin{eqnarray}
\Delta \widetilde C (\vec k,t,t',t'')
&=&\widetilde C(\vec k,t,t'')-\widetilde C(\vec k,t,t')
\nonumber\\
&=&
\frac{T}{k^2}
\left [e^{-k^2(t-t'')}-e^{-k^2(t+t'')} - e^{-k^2(t-t')} + e^{-k^2(t+t')}
\right ],
\label{tildedc}
\\
\widetilde C_{\Phi \Phi}(\vec k,t',t'')
&=&
\widetilde C(\vec k,t'',t'')+
\widetilde C(\vec k,t',t')-2\widetilde C(\vec k,t',t'') 
\nonumber \\
&=&
\frac{T}{k^2}
\left[
2-e^{-2k^2t''}-e^{-2k^2t'} - 2e^{-k^2(t''-t')} +2 e^{-k^2 (t''+t')} 
\right ] ,
\label{tildephi}
\\
\widetilde C_{\phi \phi}(\vec k,t)
&=&
\widetilde C(\vec k,t,t)=\frac{T}{k^2}\left [1-e^{-2k^2t}\right ]
\; . 
\label{tildephis}
\end{eqnarray}
Let us notice that for $\vec k \to 0$ all these quantities converge to the value
\be
\widetilde \mu(\vec k; t,t'',t')=\widetilde C_{\Xi \Xi}(\vec k,t',t'')=
\widetilde C_{\Phi \Xi}(\vec k,t',t'')=\Delta \widetilde C (\vec k,t,t',t'')=
\widetilde C_{\Phi \Phi}(\vec k,t',t'')=2T(t''-t')=2T\delta
\label{k0mom1}
\ee
except 
$\widetilde C_{\phi \phi}$ that tends to the expression 
\be
\widetilde C_{\phi \phi}(\vec k,t)=2Tt.
\label{k0mom2}
\ee
In the limit of large $t-t''$, with the help of the expressions 
(\ref{tildemu})-
(\ref{tildephis}),
it is easy to check that, as $t$ grows, the ellipse $H=0$ shrinks. Moreover, 
while for any finite $\vec k$ this curve approaches an asymptotic finite size
as $t\to \infty$, for $\vec k=0$ the ellipses shrinks to zero. This is due to the
large-$t$ divergence of $\widetilde C_{\phi \phi}$, Eq.~(\ref{k0mom2}).
Hence we conclude that, as $\vec k$ is varied inside the integral in Eq.~(\ref{eqforh}),
the most severe constraint $H(\vec k,\vec x,\vec x't,t',t'',z^*_C,z^*_\mu)>0$ 
is provided by the zero momentum modes, for large $t$. We shortly denote
with $H_0(t,t',t'',z^*_C,z^*_\mu)$ the value of the function $H$ for $k=0$, namely 
$H_0(t,t',t'',z^*_C,z^*_\mu)=H(\vec k=0,\vec x,\vec x'; t,t',t'',z^*_C,z^*_\mu)$, and indicate with $\overline z_C$ and $\overline z_\mu$ the 
values of $z^*_C$ and $z^*_\mu$ for which the constraint is satisfied $H_0(t,t',t'',\overline z_C,\overline z_\mu)=0$. 
Let us now go back to the saddle 
point equations (\ref{saddle}). Since the fluctuating quantities $\widehat d_C$ and  $\widehat d_\mu$ appear 
explicitly on the left-hand-side, while $z^*_c$ 
and $z^*_\mu$ are involved 
into a complicated function on the rhs it is easier to consider the latter 
as {\it independent} variables, trying to find the values of   $\widehat d_C$ and  
$\widehat d_\mu$ for any given couple $z^*_c$ and $z^*_\mu$.
Approaching the constraint $H_0=0$, since $\overline z_C$ and $\overline z_\mu$ are finite 
and the denominators on the rhs vanish at $\vec k=0$, the integrands diverge.
Since the first small-$\vec k$ corrections to the $\vec k=0$ results (\ref{k0mom1}) and (\ref{k0mom2}) are of order $k^2$,
the integral diverges for $d\le 2$ and converges otherwise.

In $d\le 2$ the saddle point equations (\ref{saddle}) imply that, on approaching the manifold $H_0=0$,  
$\widehat d_C$ and $\widehat d_\mu$ must diverge as well. Reverting the argument,
for $\widehat d_C$ and $\widehat d_\mu$ large enough (positive or negative),
the saddle point solutions $z^*_c,z^*_\mu$ take nearly constant values 
$z^*_c\simeq \overline z_C$, $z^*_\mu\simeq \overline z_\mu$. Let us recall now
that for $t\to \infty$ the size of the constraining ellipse $H_0=0$ vanishes, thus implying 
also $\overline z_C \to 0$ and $\overline z_\mu \to 0$. Hence, in this
large time limit the solution $z^*_C,z^*_\mu$ to Eqs.~(\ref{saddle}) is approaching
the value $\overline z_C $ and $\overline z_\mu $ in an ever increasing range
of $\widehat d_C, \widehat d_\mu$ which is moving closer and closer
to the average values $d_C,d_\mu$. In this range, the integrals in Eqs.~(\ref{saddle}),
and hence all the physics of the problem,
are dominated by the $\vec k=0$ behavior of the momentum space correlators,
Eqs.~(\ref{k0mom1}) and (\ref{k0mom2}), make the joint probability~(\ref{aftsad})
symmetric under the exchange 
$\overline z_C\widehat d_C \leftrightarrow \overline z_\mu\widehat d_\mu$,
or, equivalently 
$\overline z_C\Delta \widetilde C \leftrightarrow 2T\overline z_\mu \widetilde \mu$.
This is an analogue  situation to the one encountered in the 
simpler case of the random walk, but now this property is only obeyed
asymptotically for large $t-t''$. Accordingly, one concludes that 
$\Delta \widehat C$ and $2T\widehat \mu$ are equally distributed
in a range ever increasing with $t-t''$,
in any dimension $d\le 2$ and for any choice of $(\vec x, \vec x';t,t',t'')$
provided $t-t''$ is large. 
This suggests that all the composite operators of Eq.~(\ref{momder}) 
with the same value of $n$ [with $(\vec x_i,\vec x'_i;t_i,t'_i)=(\vec x, \vec x';t,t')$ 
$\forall i$] have the same spatio-temporal scalings. 

Let us consider now the case $d>2$. Here
the integrals on the rhs of Eqs.~(\ref{saddle}) remain finite
as the manifold $H_0=0$ is approached. This implies that, upon increasing 
the absolute value of $\widehat d_C, \widehat d_\mu$ up to certain finite values, a frontier 
$F$ is met, where the limiting saddle point solutions
$\overline z_C, \overline z_\mu$ are reached. Outside $F$ 
there is no solution to Eqs.~(\ref{saddle}) in this form. This signals that 
the large-$V$ limit, that we have taken from the beginning by replacing 
sums over momenta with intergrals, namely 
$V^{-1} \sum _{\vec k} \to \int d\vec k/(2\pi )^d$, must be reconsidered
more carefully. 
Singling out the largest contribution to the integrals, which comes from the $k=0$ components, in place of Eqs.~(\ref{saddle}) one obtains
\begin{eqnarray}
\left \{
\begin{array}{rcl}
\widehat d_C
&=&
\displaystyle
\frac{T\delta}{2V} \frac{1-\frac{T}{2}[(\delta-2t)z^*_\mu+(\delta-4t)z^*_C]}
{H_0(t,t',t'',z^*_C,z^*_\mu)}
\nonumber\\
&&
\displaystyle
+
\frac{1}{2V}\sum_{\vec k}{}^{'} \frac{\frac{1}{2}e^{i\vec k|\vec x-\vec x'|}\Delta \widetilde C
-\frac{1}{8}e^{2i\vec k|\vec x-\vec x'|}
[2T\widetilde \mu \Delta \widetilde C -2\widetilde C _{\phi \phi} 
\widetilde C_{\Phi \Xi}]z^*_\mu 
-\frac{1}{8}e^{2i\vec k|\vec x-\vec x'|}
[(\Delta \widetilde C)^2-4\widetilde C_{\phi \phi} \widetilde C_{\Phi \Phi}]z^*_C}
{H(\vec k,\vec x,\vec x'; t,t',t'',z^*_C,z^*_\mu)}
\; , 
\\
\widehat d_\mu
&=&
\displaystyle
\frac{T\delta}{2V} \frac{1-\frac{T}{2}[(\delta-2t)z^*_C+(\delta-4t)z^*_\mu]}
{H_0(t,t',t'',z^*_C,z^*_\mu)}
\nonumber\\
&&
\displaystyle
+
\frac{1}{2V}\sum_{\vec k}{}^{'} \frac{\frac{1}{2}e^{i\vec k|\vec x-\vec x'|}2T \widetilde \mu
-\frac{1}{8}e^{2i\vec k|\vec x-\vec x'|}
[2T\widetilde \mu \Delta \widetilde C -2\widetilde C _{\phi \phi} 
\widetilde C_{\Phi \Xi}]z^*_C
-\frac{1}{8}e^{2i\vec k|\vec x-\vec x'|}
[(2T \widetilde \mu)^2-4\widetilde C_{\phi \phi} \widetilde C_{\Xi \Xi}]z^*_\mu}
{H(\vec k,\vec x,\vec x'; t,t',t'',z^*_C,z^*_\mu)}
\; ,
\end{array}
\right .
\label{saddlek0}
\end{eqnarray}
where the first term on the rhs is the $k=0$ term and $\sum _{\vec k}'$
denotes the sum over all the wavevector excluding $k=0$. Inside $F$ the first term
is negligible and taking the large-$V$ limit one recovers Eqs.~(\ref{saddle}) which
admit a solution. Outside $F$,  requiring
the existence of the solution, in the large-$V$ limit the first terms must equal  
$\widehat d_C- \overline d_C$ and $\widehat d_\mu- \overline d_\mu$, respectively,
while the sums $\sum _{\vec k}^{'}$ transform back to the converging integrals
of Eqs.~(\ref{saddle}). The saddle point solution outside $F$ is therefore sticked
to the limiting value $\overline z_C, \overline z_\mu$. 
Interestingly enough, this implies that $P(\Delta \widehat C,\widehat \mu)$ has a singular point (a discontinuity of a derivative) on $F$, a feature already observed in other non-equilibrium probability distributions \cite{crisrit}.
For values of $\widehat d_C,\widehat d_\mu$ well outside $F$, namely for large
fluctuations, the contributions provided by the $k=0$ momentum dominate in 
Eqs.~(\ref{saddlek0}). If we reason now as done for the case $d\le 2$ we find the 
same conclusion as regards the distribution of fluctuations (namely $\Delta \widehat C$ and $2T\widehat \mu$ are equally distributed) and the scalings of the momenta.

Let us emphasize that the scaling properties found in the large $t-t''$ sector are
fully determined by the $\vec k=0$ slow momentum in any dimension $d$.

\subsection{Composite operators}

The fluctuating two-point operator the average of which is the linear response 
(in real space) is $\hat R_i = \phi(\vec x_i,t_i) \xi(\vec x'_i,t_i')/(2T)$. In consequence,
the higher order correlation ${\cal D}^{(n,m)}$ is given by 
\begin{eqnarray}
&&
{\cal D}^{(n,m)} = 
\langle \prod_{i=m+1}^n \phi(\vec x_i,t_i) \dot \phi(\vec x_i,t'_i) \prod_{j=1}^m \phi(\vec x_j,t_j) 
\xi(\vec x'_j,t_j')/(2T) \ \rangle 
\end{eqnarray}
This is a product of $n$ Gaussian fields, more precisely, $n-m$ factors $\phi\dot \phi$ and $m$ factors $\phi\xi$. 
Wick's theorem allows us to factor this product into products of two-field averages of the form $\langle \phi \phi\rangle$, 
$\langle \phi\dot\phi\rangle$, $\langle\dot\phi \dot\phi\rangle$, $\langle\phi\xi\rangle$ and $\langle\dot\phi \xi\rangle$. 
These are simply $C$, $\partial C$, $\partial \partial C$, $2TR$ and $2T \partial R$ (where $\partial$ indicates a 
time-derivative and one has to be careful about which is the time it is acting upon). It is not difficult to see
that in the regime of largely separated times such that all ratios are order one
\begin{equation}
1\ll \Lambda^2(t_k-t_l)  \qquad \mbox{and} \qquad  \frac{t_k}{t_l} = {\mathcal O}(1)
\; , 
\end{equation}
and for short distances
\begin{equation}
|\vec x_k -\vec x_l|^2 \ll (t_k - t_l)
\; ,
\end{equation} 
so as to make the expressions simpler,
the correlation ${\cal D}^{(n,m)}$ scales as
\begin{eqnarray}
{\cal D}^{(n,m)} &\simeq& (2T)^{n-m}
\left[ 
\prod_{i=m+1}^n {t'_i}^{-d/2} f_{\partial C}\left( \frac{t_i'}{t_i} \right) \prod_{j=1}^m {t_j'}^{-d/2} f_R
\left(\frac{t_j'}{t_j} \right) \right.
\nonumber\\
&& 
+
f_C\left( \frac{t_{m+2}'}{t_{m+1}} \right) t'^{-d}_{m+2}\, 
f_{\partial\partial C} \left(\frac{t'_{m+2}}{t'_{m+1}} \right) \prod_{i=m+3}^n
{t'_i}^{-d/2} f_{\partial C}\left( \frac{t_i'}{t_i} \right) \prod_{j=1}^m {t_j'}^{-d/2} f_R\left(\frac{t_j'}{t_j} \right) 
\nonumber\\
&&
+ \dots
\Big]
\nonumber\\
&\propto&
(2T)^{n-m} \prod_{k=1}^n {t'_k}^{-d/2} 
\nonumber
\end{eqnarray}
where the proportionality is given by a function of all ratios of times. 
This implies that the 
generalized FD ratio is also finite
\begin{equation}
X^{(n;m,m')} = \frac{{\cal D}^{(n,m)}}{{\cal D}^{(n,m')}} \propto \left( \frac{1}{2} \right)^{m-m'}  
\end{equation}
where the proportionality is also given here by a function of order one that depends on all ratios 
of the times involved in the ${\cal D}$'s. This result is akin to the one in Eq.~(\ref{xrw}) 
that was obtained for the random walk.

\section{The bidimensional clock and $XY$ models}
\label{sec:clock}

The $p$-state clock model is defined by the Hamiltonian
\be
H[\sigma]=-J\sum_{\langle ij\rangle}\vec \sigma _i\vec \sigma _j=
-J\sum_{\langle ij\rangle} \cos(\phi _i-\phi _j)
\; , 
\ee
where $\vec \sigma _i\equiv (\sigma_1^{(1)},\sigma_i^{(2)} )$ 
is a two-components unit vector spin pointing along one of the $p$ directions 
$\arctan (\sigma_i^{(2)}/\sigma_i^{(1)} )\equiv \phi _i  = 2\pi n_i /p$
with $n_i \in {1, 2, ..., p}$. $\langle ij \rangle$ denotes nearest-neighbor sites $i, j$ 
on a, in our case, square lattice in spatial dimension
$d = 2$. This spin system is equivalent to the Ising model for 
$p = 2$ and to the XY model for $p\to \infty$.
For $p \le 4$ the clock model has a critical point separating a 
disordered from an ordered phase at $T = T_1$. For $p \ge 5$
there exist two transition temperatures $T_1$ and $T_2 > T_1$ 
\cite{ref7noiclock}. For $T < T_1$ the system is ferromagnetic, 
and for $T > T_2$ it is in a paramagnetic phase. Between these two 
temperatures, for $T_1 < T < T_2$, a KT phase 
\cite{ref8noiclock} exists where the correlation function behaves 
as $G_{eq} (r) \sim | r |^{-\eta(T )}$ with the anomalous dimension 
$\eta (T )$ continuously depending on the temperature. Both 
transitions are of the KT type, namely the correlation length 
diverges exponentially as $T_1$ or $T_2$ are approached from the 
ferromagnetic or the paramagnetic phase, respectively.
The lower transition temperature goes to zero  \cite{ref7noiclock,ref9noiclock}
(approximately as $T_1\sim p^{-2}$) as $p$ grows large,
whereas $T_2$ remains finite.

In the following, dynamics are introduced by randomly 
choosing a spin and updating it with the Metropolis transition rate
\be
w([\sigma]\to[\sigma'])=\min[1,\exp (-\Delta E/T)] \; ,
\label{eq:MC-rule}
\ee
where $[\sigma]$ and $[\sigma']$ are the spin configurations before and 
after the move, and $\Delta E=H[\sigma ']-H[\sigma]$. In the limit $p\to\infty$, 
in which the angle becomes a continuous variable, Langevin dynamics 
can also be used. We give our conventions  for these dynamics in Sec.~\ref{sec:heating-2dXY} 
where we introduce the spin-wave approximation of the $2d$ XY model
and we develop our analytic results. The numerical ones of Sec.~\ref{numeric} follow 
rule~(\ref{eq:MC-rule}). 

We will consider the non-equilibrium process in which a system
of infinite size, initially (at $t=0$) in equilibrium at 
temperature $T_i$, evolves for $t>0$ in contact with a thermal bath 
at a new temperature $T$. 
Various aspects of the kinetics of the model after such a  thermal jump
have been considered in \cite{altriclock,noiclock}.
In the present paper, we will always restrict
the discussion to $p\geq 5$ and final temperatures $T$ in the KT phase, and
we will consider the heating process starting from $T_i=0$ 
or the quenching protocol where $T_i=\infty$. 
The heating case with $p\to\infty$ can be treated analytically
in the  spin-wave approximation. This will be the subject
of the next section. The results of this approach will prove to
be useful also for other cases, namely quenched or heated systems with arbitrary
$p\geq 5$, that will be studied numerically in Sec.~\ref{numeric}. 

\subsection{Heating from $T=0$ in the $p\to\infty$ model}
\label{sec:heating-2dXY}

In this Section we study analytically the Langevin dynamics of the clock model
in the limit $p\to\infty$, i.e. the $2d$ XY model. We first recall the known behavior of the averaged 
two-point and two-time correlation and linear response functions. Next we present our results for all 
moments of the fluctuating quantities the averages of which yield the usual correlation and linear 
response.

\subsubsection{The averaged correlation and linear response}
\label{subsec:averaged-quantities}

In the spin-wave approximation the free energy functional 
reads~\cite{ref26mauro}
\be
F[\phi]=\frac{\rho(T)}{2}\int d\vec x \
\left [\nabla \phi (\vec x)\right ] ^2,
\label{free}
\ee
where $\rho(T)$ is the spin-wave stiffness. The dynamics are
described by the Langevin equation
\be
\frac{\partial \phi (\vec x,t)}{\partial t}=
-\frac{\delta F[\phi]}{\delta \phi (\vec x,t)}
+\xi (\vec x,t) \; ,
\label{lang}
\ee
where the thermal noise obeys $\langle \xi (\vec x,t)\rangle =0$
and $\langle \xi (\vec x,t)\xi (\vec x',t')\rangle =
4\pi \eta(T)\rho (T)$ $\delta(\vec x-\vec x')\delta(\vec t-\vec t')$,
and the relation $2\pi \eta (T)\rho(T)=T$ holds \cite{ref19mauro}.
To ease the notation we set $\rho(T)=1$; 
indeed, it is clear from Eq.~(\ref{lang}) that the actual behavior
with $\rho(T)\neq 1$ can be recovered at the end of the calculation
by a re-definition of $\eta(T)$ and a trivial time re-scaling.
Similarly, we set $k_B=1$. With these propositions Eq.~(\ref{lang}) 
is equal to Eq.~(\ref{eq:Langevin-scalar}), so that we can borrow the results
of Sec.~\ref{sec:scalar} whenever it will be needed to infer the properties
of the actual XY system. In order to avoid confusion between the two models,
quantities relative to the scalar field will be denoted with an index $_\phi$
(e.g. $C_\phi$ and $R_\phi$ will be the correlation and response of the scalar field,
already given in Eqs.~(\ref{eq:gamma-scalar}) and (\ref{eq:bare-R0-scalar})).

From the knowledge of the angle dynamics the spin correlation
\be
C(r;t,t')=\langle \cos[\phi(\vec x,t)-\phi(\vec x',t')]\rangle
\ee
can be readily evaluated. The spin linear response function
\be
R(r;t,t')\equiv \sum _{\beta =1}^2 R^{(\beta)}(r;t,t')\equiv 
\sum _{\beta =1}^2 \left.
\frac{\delta \langle \sigma ^{(\beta)}(\vec x,t)\rangle_{\vec h}}
{\delta h^{(\beta)}(\vec x', t')}  
\right|_{h=0} ,
\ee
where the vector $\vec h\equiv(h^{(1)},h^{(2)})$
is the perturbation conjugated to $\vec \sigma $ [i.e. by adding 
$-\rho(T)/2 \int d\vec x \ \vec \sigma(\vec x) \vec h(\vec x,t)$ 
to the free-energy] can be obtained from
\be
R(r;t,t')=\frac{1}{2T}\langle \xi(\vec x',t')\sin[\phi(\vec x,t)-\phi(\vec x',t')]\rangle 
\; .
\ee 
The averaged quantities $C(r;t,t')$ and $R(r;t,t')$ and their relation have been
studied in~\cite{cuglia94,Behose01,Abriet03,Lei07}. We will recover these functions as special cases in Sec.~\ref{scalcompx}.

\subsubsection{Composite operators} \label{scalnthmom}

In this paper we are interested in the more general
problem of the pdf of the fluctuations of
the correlation and linear response. In this case it is convenient to construct the
pdf by evaluating all the moments. Let us start by defining the fluctuating quantities 
we are interested in as
\begin{eqnarray}
&&
\widehat C_i = \widehat C(\vec x_i,\vec x'_i;t_i,t'_i)
=\cos\delta_i(\vec x_i, \vec x'_i; t_i,t'_i) 
\; , 
\label{flucC}
\\
&&
\widehat R_i=\widehat R(\vec x_i,\vec x'_i;t_i,t'_i) =
\frac{1}{2T}\xi(\vec x'_i,t'_i)\sin\delta _i(\vec x_i, \vec x'_i; t_i,t'_i) 
\; ,
\end{eqnarray}
where $\delta _i(\vec x_i,\vec x'_i;t_i,t'_i) \equiv \phi(x_i,t_i)-\phi(x'_i,t'_i)$ and,
as before, 
$x_i$ and $x'_i$ are two generic points in space and
$t_i$ and $t'_i$ two generic times (still with $t_i \geq t_i'$). In the following we will not write the explicit 
space and time dependences in $\delta_i$ to simplify the notation. For the same reason
we will also set $T=1$. Starting from these
definitions one can build averages of composite fields of the form Eqs.~(\ref{momder}) and (\ref{momC}).
These can be computed with the help of the {\it generator} ${\cal C}_\lambda^{(n,0)}$,
which is obtained by 
adding the extra angle
$\alpha _{\lambda_i}$ to $\delta _i$, namely replacing  $\delta _i$
with $\delta _{i,\lambda}=\delta _i +\alpha_{\lambda _i}$
in Eq.~(\ref{flucC}). More precisely, we define
\be
{\cal C}_\lambda^{(n,0)} \equiv \langle \ \prod_{i=1}^{n} \widehat C_{i,\lambda} \ \rangle
\label{gen}
\ee
where
\be
\widehat C_{i,\lambda}
=\cos\delta _{i,\lambda}
\ee
We choose the extra angle such that $\alpha _{\lambda _i=0}=0$. 
Then one trivially recovers the quantity in Eq.~(\ref{momC}), for the special case 
$m=0$, as $\left . {\cal C}^{(n,0)}={\cal C}_\lambda^{(n,0)}\right \vert _{\{ \lambda \} =0}$,
where $\{\lambda \}=0$ means $\lambda _i=0$ $\forall i$.
Defining $\partial {\cal C}_{\{\lambda \}=0}^{(n,0)} \equiv
\left . \partial _{\lambda_1} ... \partial _{\lambda_n}
{\cal C}_\lambda^{(n,0)}\right \vert _{\{ \lambda \} =0}$, 
where as before $\partial _\lambda=\partial /\partial \lambda$ means 
the derivative with respect to a generic argument $\lambda$,
one has
\be
\partial {\cal C}_{\{\lambda\}=0}^{(n,0)}=
\langle \ \prod_{i=1}^{n} 
\partial \alpha _{\lambda _i} 
\sin\delta _i \ \rangle
\; ,
\ee
where $\partial \alpha _{\lambda _i}=
\partial _{\lambda _i}\alpha _{\lambda _i}\vert _{\lambda_i=0} $.
This quantity provides all the composite fields in Eq.~(\ref{momder})
when the choices $\alpha_{\lambda_i}^{(\partial C)}$ for $i=1,\dots, m$,
and  $\alpha_{\lambda_i}^{(R)}$ for $i=m+1,\dots, n$, are respectively made, with 
\begin{eqnarray}
\partial \alpha ^{(\partial C)}_{\lambda _i}
&=&
-\partial _{t'_i}\phi (\vec x'_i,t'_i)
\; ,
\label{propdC}
\\
\partial \alpha^{(R)} _{\lambda _i}
&=&
-\frac{\xi (\vec x'_i,t'_i)}{2T}
\; . 
\label{propR}
\end{eqnarray}
The computation of the generator (\ref{gen}) (see App.~\ref{app1})
yields 
\be
{\cal C}_\lambda^{(n,0)}=2^{-n}\sum _{\{s_i=\pm 1\}} e^{-\frac{1}{2}\langle {\cal S}_{\lambda}^2 \rangle }
\; ,
\label{c2}
\ee
where ${\cal S}_\lambda=\sum_{i=1}^ns_i\delta_{i,\lambda}$ and $s_i$ are auxiliary Ising variables 
introduced for convenience. In the long times sector  in which $2\Lambda^2(t+t')\gg 1$ and 
$2\Lambda^2(t-t')\gg 1$, so that we can use the limiting behavior of $C_\phi$ and $R_\phi$ 
discussed in Sec.~\ref{subsec:averaged-quantities},
one obtains
\be
\partial {\cal C}^{(n,0)}_{\{\lambda \}=0}=2^{-n}\sum _{\{s_i=\pm 1\}}\left (\prod_{i=1}^{n}s_i\right )
P^{(n)}(\{\langle {\cal S} \partial \alpha _{\lambda _i}\rangle \})
\
e^{-\frac{1}{2}\langle {\cal S}^2 \rangle }
\; ,
\label{dergen}
\ee 
where ${\cal S}= {\cal S}_{\lambda=0}$ and 
the polynomial $P^{(n)}$ is defined by the recursive relation given in Eq.~(\ref{recursive}). 
Given the form of ${\cal S}$ and Eqs.~(\ref{propdC}) and (\ref{propR})
these quantities are all expressed in terms of the angle correlation
and responses, $C_\phi$ and $R_\phi$, and can, therefore, be explicitly
calculated. 

Let us come now to the composite operators ${\cal D}^{(n,m)}$ defined 
in Eq.~(\ref{momder}), which
can all be obtained from Eq.~(\ref{dergen}). By comparing two different moments
with the same $n$ but different $m$ (say $m$ and $m'>m$) the difference is only
due to the replacement 
$\partial \alpha_{\lambda _i}^{(R)}\to \partial \alpha_{\lambda _i}^{(\partial C)}$ 
$\forall i=m+1,\dots,m'$ in the polynomial $P^{(n)}$, which in turn amounts
to the substitution of the functions $\partial _{t'}C_\phi (\vert x_i-x'_i\vert;t_i,t'_i)$
with $R_\phi(\vert \vec x_i-\vec x'_i\vert;t_i,t'_i)$ $\forall i=m+1,\dots,m'$.
For $|\vec x_i -\vec x'_i|\ll 2\Lambda ^2 (t_i-t'_i)$ $\forall i$ 
these quantities are proportional, according
to Eq.~(\ref{xtheta}). Therefore one concludes that all the composite operators
(and hence all the moments) with equal $n$ scale in the same way, and
the generalized FD ratios (\ref{xmom-def})
depend only on the ratios $y_i=t_i/t'_i$ and
are independent of the distances $\vert \vec x_i-\vec x'_i\vert$. 
Furthermore, according to Eq.~(\ref{xthetalim}), for $y_i\to \infty$ $\forall i$
the limiting value    
\be
X^{(n;m,m')}_{\infty} \equiv \lim _{t/t'\to \infty} X^{(n;m,m')}
\label{xmomlim}
\ee          
is finite and independent of the spatial arguments.

\subsubsection{Scaling of composite operators and their FD
ratio for $n=1,2$} \label{scalcompx}

As a concrete example we now explicitate the expressions obtained in Sec.~\ref{scalnthmom} for
the simplest cases with $n=1,2$. The generalization to generic values
of $n$ is straightforward.  Starting with
the case $n=1$,
from Eq.~(\ref{c2}) one immediately obtains
\be
{\cal C}^{(1,0)}=C(r;t,t')=e^{-\frac{1}{2}\left [\phi (\vec x,t) -\phi (\vec x',t')\right]^2}=
e^{-\frac{1}{2}[C_\phi (0;t,t)+C_\phi (0;t',t')-2C_\phi (r;t,t')]}
\ee
where we dropped the sub-index $1$, namely we set $x_1=x$ and similarly for the
other variables. For the computation of the ${\cal D}$'s we enforce Eq.~(\ref{dergen})
using, according to Eq.~(\ref{recursive}),
\be 
P^{(1)}(\{\langle {\cal S} \partial \alpha _{\lambda _i}\rangle \})=
-s_1\langle \delta_1 \partial \alpha _{\lambda _1}\rangle
\; .
\ee 
Hence one arrives at 
\be
\partial {\cal C}^{(1,0)}_{\{\lambda\}=0}=\langle[\phi (\vec x;t)-\phi (\vec x';t')] 
\partial \alpha _\lambda\rangle C(r;t,t')
\; ,
\ee
and from here
\be
{\cal D}^{(1,0)}=\partial_{t'}C(r;t,t')=
[\partial _{t'}C_\phi (r;t,t')-\partial _{t'}C_\phi (0;t',t')]\,C(r;t,t')
\ee
and
\be
{\cal D}^{(1,1)}=R(r;t,t')=[R_\phi (r;t,t')-R_\phi (0;t',t')]\,C(r;t,t')
\; . 
\ee
Assuming $R_\phi(r;t,t)=\partial C_\phi(r;t,t)\equiv 0$,
for the FD ratio $X(r;t,t')=X^{(1;1,0)}(r;t,t')$ one finds
\be
X(r;t,t')=X_\phi(r;t,t')
\; ,
\ee
showing that the FD ratio of the XY model is the same as
that of the scalar field.

Let us now consider the case with $n=2$.  Proceeding analogously to the case $n=1$,
from Eq.~(\ref{c2}) one has 
\be
{\cal C}^{(2,0)}=\frac{1}{2}\left [{\cal C}^{(2,0)}_{+}+{\cal C}^{(2,0)}_{-}\right ]
\label{c2_1}
\ee
where
\begin{eqnarray}
{\cal C}^{(2,0)}_{\pm}
&=& 
\exp\left\{
-\frac{1}{2}[C_\phi(0;t_1,t_1)+C_\phi(0;t'_1,t'_1)+C_\phi(0;t_2,t_2)+C_\phi(0;t'_2,t'_2)
\right.
\nonumber\\
&&
\left.
\qquad
-2C_\phi(|\vec x_1-\vec x'_1|;t_1,t'_1)-2C_\phi(|\vec x_2-\vec x'_2|;t_2,t'_2)]
\right\}
 \nonumber \\
&&
\times \exp \left\{ 
\pm 2[C_\phi(|\vec x_1-\vec x_2|;t_1,t_2)
-C_\phi(|\vec x_1-\vec x'_2|;t_1,t'_2) 
\right.
\nonumber\\
&&
\left.
\qquad\qquad
-C_\phi(|\vec x'_1-\vec x_2|;t_2,t'_1)
+C_\phi(|\vec x'_1-\vec x'_2|;t'_1,t'_2)]
\right\} 
\; . 
\label{c2pm}
\end{eqnarray}
The second order recursive polynomial reads
\be
P^{(2)}(\{\langle {\cal S} \partial \alpha _{\lambda _i}\rangle \})=
\langle {\cal S}\partial \alpha _{\lambda _1}\rangle 
\langle {\cal S}\partial \alpha _{\lambda _1}\rangle
-s_2\langle \partial \alpha _{\lambda _1} \partial \alpha _{\lambda _2}\rangle
\ee 
where, using the terminology of App.~\ref{app1}, the two terms on
the rhs are of type $f$ and $\overline f$, respectively.
Neglecting the latter, since we show in App.~\ref{app2}  that it is  subdominant in the large time sector,
we arrive at
\begin{eqnarray}
\partial {\cal C}^{(2,0)}_{\{\lambda\}=0}
&=&
\frac{1}{2}\left [
\langle(\delta_1+\delta_2)\partial \alpha _{\lambda _1}\rangle 
\langle(\delta_1+\delta_2)\partial \alpha _{\lambda _2}\rangle
{\cal C}^{(2,0)}_{+}
\right.
\nonumber\\
&& 
\left. 
\qquad +
\langle(\delta_1-\delta_2)\partial \alpha _{\lambda _1}\rangle
\langle (\delta_1-\delta_2)\partial \alpha _{\lambda _2}\rangle
{\cal C}^{(2,0)}_{-}\right ]
\end{eqnarray}
or, more explicitly,
\be
{\cal D}^{(2,m)}=\frac{1}{2}\left [A_+^{(2,m)}{\cal C}^{(2,0)}_++A_-^{(2,m)}{\cal C}^{(2,0)}_-\right ]
\ee
with
\begin{eqnarray}
A_{\pm}^{(2,0)}
&=&
\left [\partial _{t'_1}C_\phi(|\vec x_1-\vec x'_1|;t_1,t'_1)-
\partial _{t'_1}C_\phi(0;t'_1,t'_1) 
\right.
\nonumber\\
&&
\qquad
\left.
\pm \partial _{t'_1}C_\phi(|\vec x_2-\vec x'_1|;t_2,t'_1)
\mp \partial _{t'_1}C_\phi(|\vec x'_2-\vec x'_1|;t'_2,t'_1)
\right]
\nonumber\\
&&
\times
 \left [\partial _{t'_2}C_\phi(|\vec x_1-\vec x'_2|;t_1,t'_2) 
-\partial _{t'_2}C_\phi(|\vec x'_1-\vec x'_2|;t'_1,t'_2)
\right.
\nonumber\\
&&
\qquad
\left.
\pm \partial _{t'_2}C_\phi(|\vec x_2-\vec x'_2|;t_2,t'_2)
\mp \partial _{t'_2}C_\phi(0;t'_2,t'_2)
\right]
\; , 
\nonumber\\
A_{\pm}^{(2,1)}
&=&
\left [\partial _{t'_1}C_\phi(|\vec x_1-\vec x'_1|;t_1,t'_1)-
\partial _{t'_1}C_\phi(0;t'_1,t'_1) 
\right.
\nonumber\\
&&
\qquad
\left.
\pm \partial _{t'_1}C_\phi(|\vec x_2-\vec x'_1|;t_2,t'_1)
\mp \partial _{t'_1}C_\phi(|\vec x'_2-\vec x'_1|;t'_2,t'_1)
\right] 
\nonumber\\
&&
\times
 \left [R_\phi(|\vec x_1-\vec x'_2|;t_1,t'_2) 
-R_\phi(|\vec x'_1-\vec x'_2|;t'_1,t'_2)
\right.
\nonumber\\
&&
\qquad
\left.
\pm R_\phi(|\vec x_2-\vec x'_2|;t_2,t'_2)
\mp R_\phi(0;t'_2,t'_2)
\right]
\; , 
\nonumber\\
A_{\pm}^{(2,2)}
&=&\left [R_\phi(|\vec x_1-\vec x'_1|;t_1,t'_1)-
R_\phi(0;t'_1,t'_1) 
\right.
\nonumber\\
&&
\qquad
\left.
\pm R_\phi(|\vec x_2-\vec x'_1|;t_2,t'_1)
\mp R_\phi(|\vec x'_2-\vec x'_1|;t'_2,t'_1)
\right] 
\nonumber\\
&&
\times 
\left [R_\phi(|\vec x_1-\vec x'_2|;t_1,t'_2) 
-R_\phi(|\vec x'_1-\vec x'_2|;t'_1,t'_2)
\right.
\nonumber\\
&&
\qquad
\left.
\pm R_\phi(|\vec x_2-\vec x'_2|;t_2,t'_2)
\mp R_\phi(0;t'_2,t'_2)\right ].
\end{eqnarray}

In the following, in order to simplify the discussion, we focus
on the case 
$\vec x'_1=\vec x_1\,, \vec x'_2=\vec x_2$ 
and $t_1=t_2=t\,,t'_1=t'_2=t'$. This choice will be adopted in 
Sec.~\ref{numeric} for the numerical computations.
Letting $r^2\equiv (\vec x_1 -\vec x_2)^2\ll 2\Lambda ^2 (t-t')$
and using the scaling form for $R_\phi$ and $\partial _{t'}C_\phi$
derived in Sec.~\ref{sec:scalar} [below Eq.~(\ref{eq:bare-R01-scalar})], 
one easily obtains
\be
X^{(n=2;m,m')}(r;t,t')=\left[ X_\phi(y) \right]^{m-m'},
\ee
where $y=t/t'$ as usual. The limiting FD ratio
is given by
\be
X^{(n=2;m,m')}_\infty \equiv \lim_{y\to\infty} X^{(n=2;m,m')}(r;t,t') =\left[ X_{\phi,\infty}(y)\right]^{m-m'}
=
\left (\frac{1}{2}\right )^{m-m'}
\; . 
\ee
Proceeding analogously, one can generalize the computation of 
$X_\infty^{(n;m,m')}$ to any value of $n,m,m'$ (but the calculation becomes
lengthy upon increasing $n$).

\subsubsection{Moments} \label{moments}

Moments can be defined from the composite operators (\ref{momder}) and (\ref{momC}) by
subtracting a suitable disconnected part. In view of the numerical
applications of Sec.~\ref{numeric} we will concentrate here on the
time-integrated quantities of Eq.~(\ref{momC}). These quantities
are less numerically demanding than the corresponding differential ones (\ref{momder}).
For the same reason we restrict to the case $n=2$.
With the choice $\vec x'_1=\vec x_1\,, \vec x'_2=\vec x_2$ 
and $t_1=t_2=t\,,t'_1=t'_2=t'$ made in the previous section,
moments can be defined as 
\be
{\cal V}^{(2,m+m')}(\vec x_1,\vec x_2;t,t')=
\langle \widehat {\cal C}^{(1,m)}(\vec x_1;t,t')
\widehat {\cal C}^{(1,m')}(\vec x_2;t,t') \rangle - 
{\cal C}^{(1,m)}(t,t') {\cal C}^{(1,m')}(t,t') 
\label{var}
\ee
where $\widehat {\cal C}$ is the fluctuating part (insides brakets $\langle \dots \rangle$)
of ${\cal C}$ in Eq.~(\ref{momC}).
In order to improve the 
statistics of the concrete numerical measurements that will be discussed in the next section, 
and to compare to similar calculations presented in \cite{noivarianze,noispinglass},
we compute the double spatially integrated quantities
\be
{\cal V}^{(2,m)}_{k=0}(t,t')=L^{-2}\int d\vec x_1 \int d\vec x_2 \,\, 
{\cal V}^{(2,m)}(\vec x_1, \vec x_2; t,t')
\label{vark0}
\ee
with $L$ the linear size of the sample.
${\cal V}^{(2,0)}$ is the quantity that is usually computed when  
dynamical heterogeneities in disordered and glassy systems are studied~\cite{c4}. 
Together with this one, the other two quantities have been studied
in different aging systems and with different techniques in~\cite{noivarianze,annibsollich,noispinglass}.

Using Eqs.~(\ref{c2_1}) and (\ref{c2pm}), ${\cal V}^{(2,0)}_{k=0}$ can be written as
\begin{eqnarray*}
&&{\cal V}^{(2,0)}_{k=0}(t,t')=\frac{1}{2}
e^{-C_\phi(0;t,t)-C_\phi(0;t',t')+2C_\phi(0;t,t')}
\nonumber \\
&&
\qquad\left \{ \int d\vec r \left [e^{2C_\phi(r;t,t)+2C_\phi(r;t',t')-4C_\phi(r;t,t')} +
e^{-2C_\phi(r;t,t)-2C_\phi(r;t',t')+4C_\phi(r;t,t')}\right ]-2\right \},
\end{eqnarray*}
where $\vec r=\vec x'-\vec x$.
Using the expressions derived for $C_\phi$ in App.~\ref{app3}
for $2\Lambda ^2(t-t')\gg 1$ we obtain the scaling form
\be
{\cal V}^{(2,0)}_{k=0}=t'^af^{(2,0)}(y)
\label{dcba}
\ee
with $a=(2-2\eta)/z$, where $\eta =T/(2\pi)$ is the equilibrium anomalous exponent
and $z=2$ is the dynamical 
exponent. This result agrees with the general behavior
\begin{equation}
{\cal V}^{(2,m)}_{k=0}=t'^{(4-d-2\eta)/z}f^{(2,m)}(y)
\label{eq:general-scaling}
\end{equation} 
expected on the basis of 
critical scaling arguments~\cite{noivarianze}, and confirmed in~\cite{annibsollich} 
in the spherical model. 
The same scaling is obeyed also by ${\cal V}^{(2,1)}_{k=0}$
and ${\cal V}^{(2,2)}_{k=0}$, with the same large-$t/t'$ behavior of the
scaling functions, since we have proven in Sec.~\ref{scalnthmom} that
all composite operators with the same $n$ scale in the same way.
Notice that the exponent $a$ is an equilibrium property,
being only determined by the equilibrium exponents $\eta$ and $z$.

\subsection{Numerical simulations} 
\label{numeric}

The analytical results of the previous section apply to a system with $p\to\infty$
heated from $T_i=0$ to a temperature $T$ in the KT phase.
The next question is how general this picture is and, in particular,
i) what is the behavior of systems evolving in a KT phase with $p<\infty$ and   
ii) which modifications arise in a quench with $T_i=\infty$ where topological defects 
are present due to the disordered initial condition.
In this section we address these questions numerically.
In order to do so we evolved systems with $p=6$ and $12$
starting from equilibrium states at $T_i=0$ and $T_i=\infty$. 
In the case of a quench in the $p=6$ case, the behavior of $C(0;t,t')$ and
$\chi (0;t,t')$ was shown~\cite{noiclock} to fit into the general
scenario expected from standard scaling arguments (apart from logarithmic
corrections due to the presence of vortices, see the discussion below). 
Here we will concentrate on the behavior of the moments ${\cal V}^{(2,0)}_{k=0}$
of Eq.~(\ref{vark0}) which, in the present on-lattice model are obtained
as ${\cal V}^{(2,0)}_{k=0}(t,t')=L^{-2}\sum _{i\neq j}{\cal V}^{(2,m)}(\vec x_i,\vec x_j;t,t')$,
where $\vec x_i$ ($\vec x_j$) is the (square) lattice coordinate of site $i$ ($j$), and
$L^2$ is the number of lattice points.
Whenever  a response function is involved this has been computed
with the extension of the FD theorem
to non-equilibrium states derived in \cite{algo}. This method
has been thoroughly applied \cite{algonum} to study different
problems for its numerical efficiency and because, being perturbation-free, 
guarantees correct results in the linear regime. 
The working temperature is chosen to be $T=0.76$, which belongs
to the KT phase both for $p=6$ and $p=12$, and the system size is $L=600$. No finite size effects
are detected with this choice, in the range of simulated times.
The data presented are averages over $2\cdot 10^3$ - $6\cdot 10^3$ 
(according to the different cases)
realizations of the thermal noise and, in the case of quenches, of the initial 
conditions. Times are given in Monte Carlo (MC) units. Whenever we plot
the moments, we include a suitable $T$ factor to make them dimensionless;
namely we always plot ${\cal V}_{k=0}^{(2,0)}$, $T{\cal V}_{k=0}^{(2,1)}$,
and $T^2{\cal V}_{k=0}^{(2,2)}$. 

\subsubsection{Heating from zero temperature}

\vspace{0.2cm}
\noindent
{\it Clock model with $p=6$}.
\vspace{0.2cm}

\begin{figure}[h]
    \centering
   \rotatebox{0}{\resizebox{.5\textwidth}{!}{\includegraphics{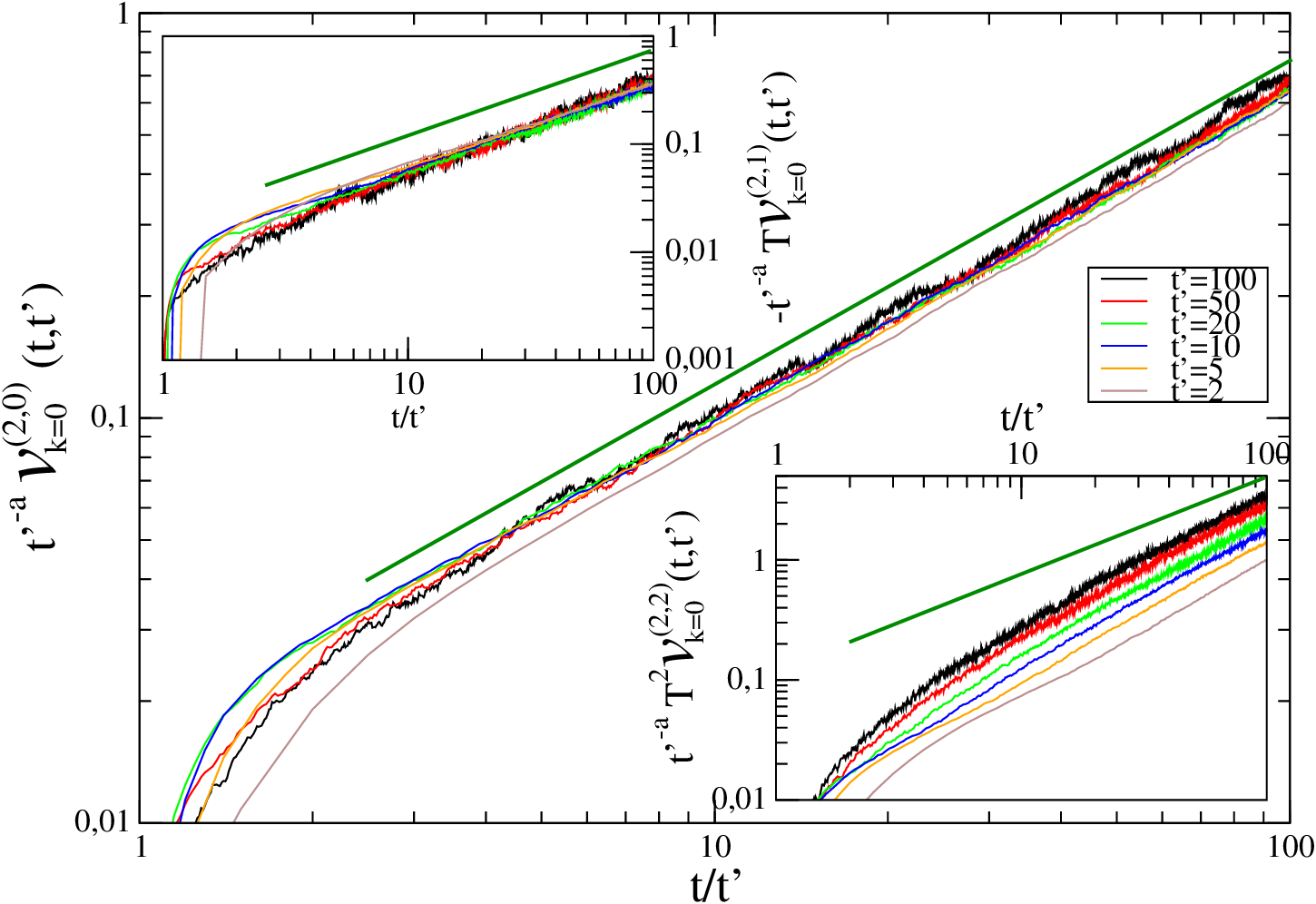}}}
      \rotatebox{0}{\resizebox{.5\textwidth}{!}{\includegraphics{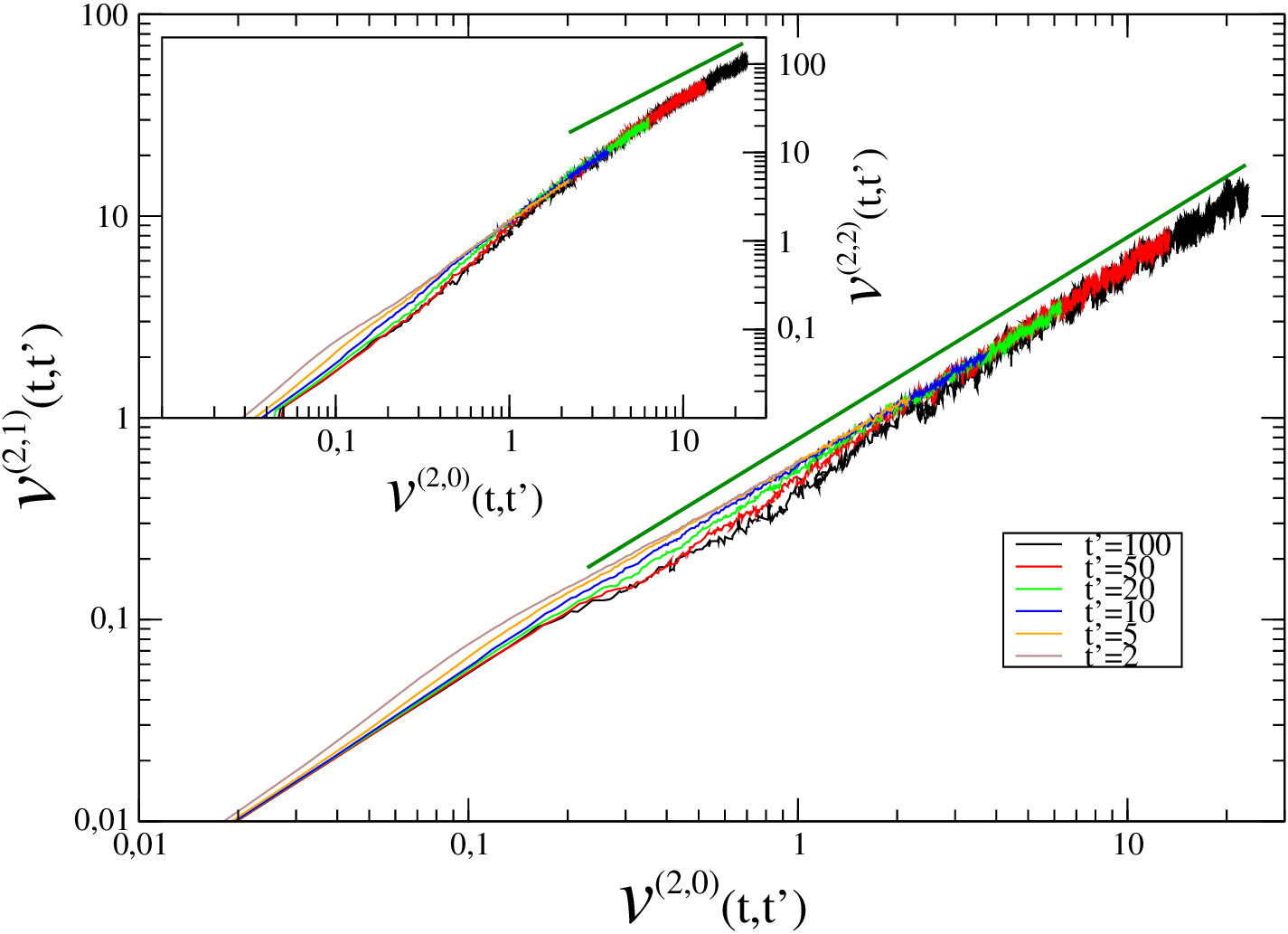}}}
   \vspace{1cm}
   \caption{(Color online.) Clock model with $p=6$ heated from $T=0$ to $T=0.76$. 
Upper panel: The re-scaled moment $t'^{-a}{\cal V}_{k=0}^{(2,0)}(t,t')$, with $a=0.76$, 
is plotted against $t/t'$ in the main part of the figure, while the same plot
for ${\cal V}_{k=0}^{(2,1)}$ (with a minus sign in order to have a positive quantity)  
and ${\cal V}_{k=0}^{(2,1)}$ is presented in the
upper-left and lower-right insets, respectively. Different choices of $t'$
correspond to different curves, see the key. The straight green bold 
line is the power law $(t/t')^{0.8}$. Lower panel: Same data as in the upper panel
but plotted in the parametric form $-{\cal V}_{k=0}^{(2,1)}(t,t')$ against 
${\cal V}_{k=0}^{(2,0)}(t,t')$ (main part) and ${\cal V}_{k=0}^{(2,2)}(t,t')$
against ${\cal V}_{k=0}^{(2,0)}(t,t')$ (inset). The straight green bold line is the linear
behavior (i.e. scaling functions are proportional).}
\label{fig-p6-heated}
\end{figure}

The numerical estimates $\eta=0.17$ and $z=2.18$ are reported in the literature 
\cite{ref9noiclock,noiclock}, from which, comparing with 
Eqs.~(\ref{dcba}) and (\ref{eq:general-scaling}) 
 one obtains $a=(4-d-2\eta)/z\simeq 0.76$.
In Fig.~\ref{fig-p6-heated} the quantities $t^{-a}{\cal V}_{k=0}^{(2,m)}$ are plotted
against $t/t'$ in the upper panel. 
One observes a nice data collapse for ${\cal V}_{k=0}^{(2,0)}$
and ${\cal V}_{k=0}^{(2,1)}$, where small corrections are only visible for the 
smallest value of $t'$ in the early regime with $t/t'$ small.
The quantity ${\cal V}_{K=0}^{(2,2)}$, instead, presents larger corrections,
and only an asymptotic trend towards a scaling collapse is observed
(the two largest value of $t'$ are almost superimposed).
The same large corrections to scaling affect also the scaling function of  
${\cal V}_{k=0}^{(2,2)}$. 
Indeed, while $f^{(2,0)}(x)$ and $f^{(2,1)}(x)$ are proportional,
as expected, and grow algebraically as $f^{(2,0)}(x)\sim f^{(2,1)} \sim x^\alpha$,
with a value of $\alpha \simeq 0.8$ (measured for $x\ge 10$), 
$f^{(2,2)}(x)$ grows with a larger effective exponent $\alpha _{eff}$ 
but, as it is more visible for the largest $t'$,
this exponent decreases as $x$ increases (for $t'=100$ one measures 
$\alpha _{eff}\simeq 0.9$ for $x\ge 70$). 
In order to better appreciate the fact that all moments scale with the same exponent,
and to test this fact in a parameter-free plot, in the lower plot of Fig.~\ref{fig-p6-heated}
we present the parametric plots of  $-{\cal V}_{k=0}^{(2,1)}$ and 
${\cal V}_{k=0}^{(2,2)}$ against ${\cal V}_{k=0}^{(2,0)}$.
We find an excellent data collapse for the different values of $t'$,
confirming once again that all the moments scale in the same way. Notice also
that data follow a linear behavior (green line) for large times, signaling that also the
the scaling functions are proportional (for ${\cal V}_{k=0}^{(2,2)}$, due to the
above-mentioned pre-asymptotic corrections, the approach to a linear behavior
is seen only for the latest data). In conclusion, our data are consistent
with an asymptotic scaling ${\cal V}_{k=0}^{(2,m)}=t'^{(4-d-2\eta)/z}f^{(2,m)}(t/t')$,
with the scaling functions increasing algebraically with an $m$-independent
exponent $\alpha$. The limiting FD ratios
$\lim _{t' \to \infty} \lim _{t\to \infty} X^{(2;m,m')}(t,t')$ are therefore finite.
We conclude  that the scaling scenario given in Eqs.~(\ref{dcba}) and (\ref{eq:general-scaling})  
and suggested by the spin-wave results
applies to this case, provided that the actual
values of $\eta$ and $z$ are taken into account. 

\vspace{1cm}

\vspace{0.2cm}
\noindent
{\it Clock model with $p=12$}
\vspace{0.2cm}

\begin{figure}[h]
    \centering
   \rotatebox{0}{\resizebox{.5\textwidth}{!}{\includegraphics{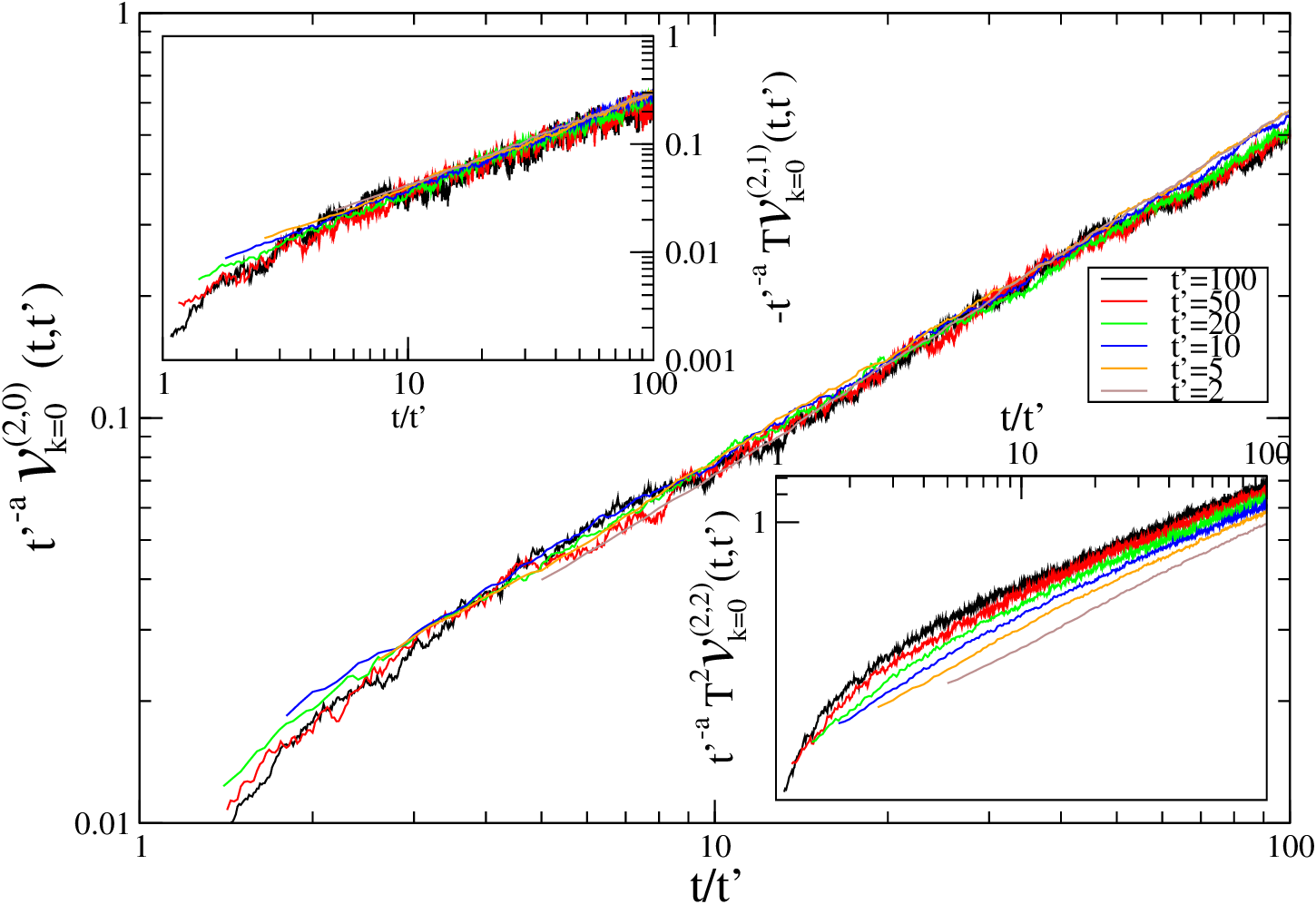}}}
      \rotatebox{0}{\resizebox{.5\textwidth}{!}{\includegraphics{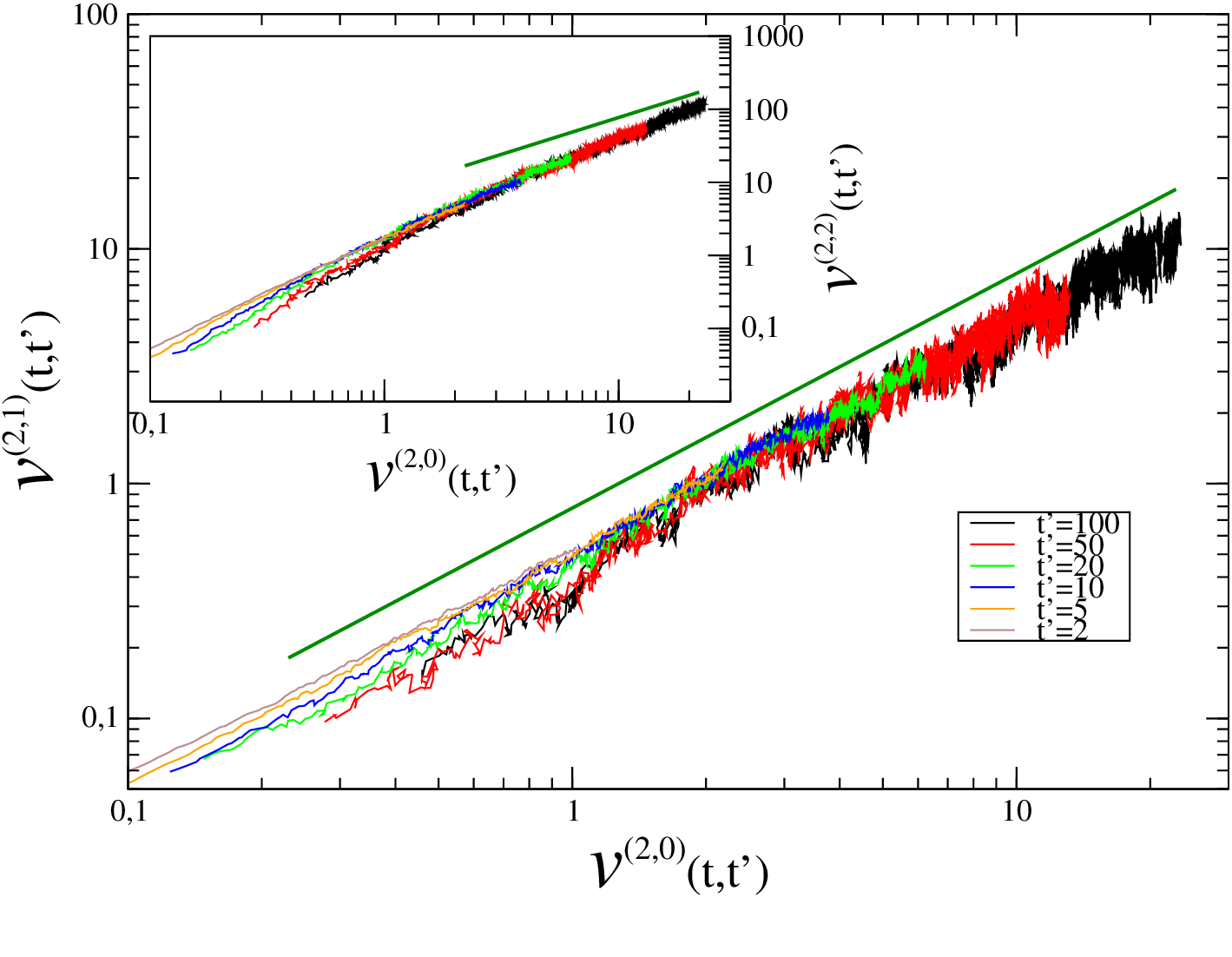}}}
   \vspace{1cm}
   \caption{(Color online.) Clock model with $p=12$ heated from $T=0$ to $T=0.76$. 
Upper panel: The re-scaled moment $t'^{-a}{\cal V}_{k=0}^{(2,0)}(t,t')$, with $a=0.83$, 
is plotted against $t/t'$ in the main part of the figure, while the same plot
for ${\cal V}_{k=0}^{(2,1)}$ (with a minus sign in order to have a positive quantity)  
and ${\cal V}_{k=0}^{(2,1)}$ is presented in the
upper-left and lower-right insets, respectively. Different choices of $t'$
correspond to different curves, see the key.
Lower panel: Same data as in the upper panel
but plotted in the parametric form $-{\cal V}_{k=0}^{(2,1)}(t,t')$ against 
${\cal V}_{k=0}^{(2,0)}(t,t')$ (main part) and ${\cal V}_{k=0}^{(2,2)}(t,t')$
against ${\cal V}_{k=0}^{(2,0)}(t,t')$ (inset). The straight green bold line is the linear
behavior (i.e. scaling functions are proportional).}
\label{fig-p12-heated}
\end{figure}

The results for $p=12$ are presented in Fig.~\ref{fig-p12-heated}.
One observes the same qualitative behavior as in the case with $p=6$.
With a value $a=0.83$ one obtains an excellent data collapse for ${\cal V}_{k=0}^{(2,0)}$
and ${\cal V}_{k=0}^{(2,1)}$, while for  ${\cal V}_{k=0}^{(2,2)}$ the collapse
is only asymptotically approached, similarly (but the collapse is somewhat better) 
to the case with $p=6$. 
Notice that this value of $a$ is basically the same as the one obtained
for $p\to\infty$, where one has $a\simeq 0.83$ since  $\eta \simeq 0.165$ and $z=2$~\cite{gupta}.
From $p=12$ onward, therefore, one does not expect to see any significant
difference with the spin-wave analytic results. The scaling functions
behave as  $f^{(2,0)}(x)\sim f^{(2,1)} \sim x^\alpha$, with $\alpha = 0.84$.
For $f^{(2,2)}$ the data are consistent with an asymptotic convergence towards
the same power-law behavior. This picture is confirmed by the parametric plots 
of  $-{\cal V}_{k=0}^{(2,1)}$ and 
${\cal V}_{k=0}^{(2,2)}$ against ${\cal V}_{k=0}^{(2,0)}$
presented in the lower panel of Fig. \ref{fig-p12-heated}.

\subsubsection{Quench from infinite temperature} \label{clockquench}

When quenching from $T_i\to\infty$ the presence of vortics makes the dynamics
quite different from the one observed in the heating case~\cite{vortici}. It is therefore interesting to 
see how the scenario provided by the spin-wave approximation 
may or may not  be modified in this case. In the following we present the results of
simulations of the same system considered insofar, but with initial
conditions extracted from an $T_i\to\infty$ equilibrium ensemble.
We only discuss the case with $p=6$ since we found very 
similar results for $p=12$.

\begin{figure}[h]
    \centering
   \rotatebox{0}{\resizebox{.5\textwidth}{!}{\includegraphics{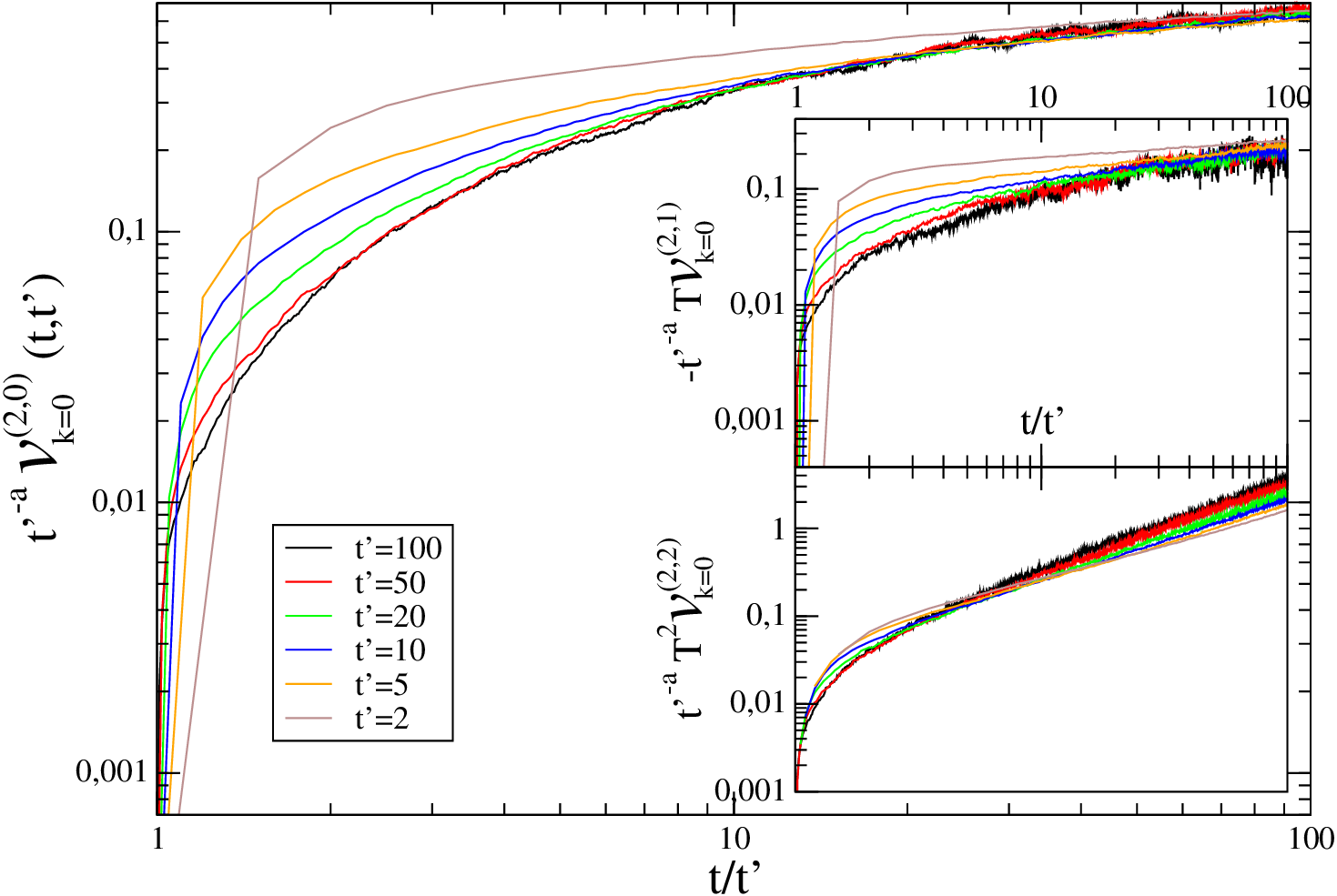}}}
      \rotatebox{0}{\resizebox{.5\textwidth}{!}{\includegraphics{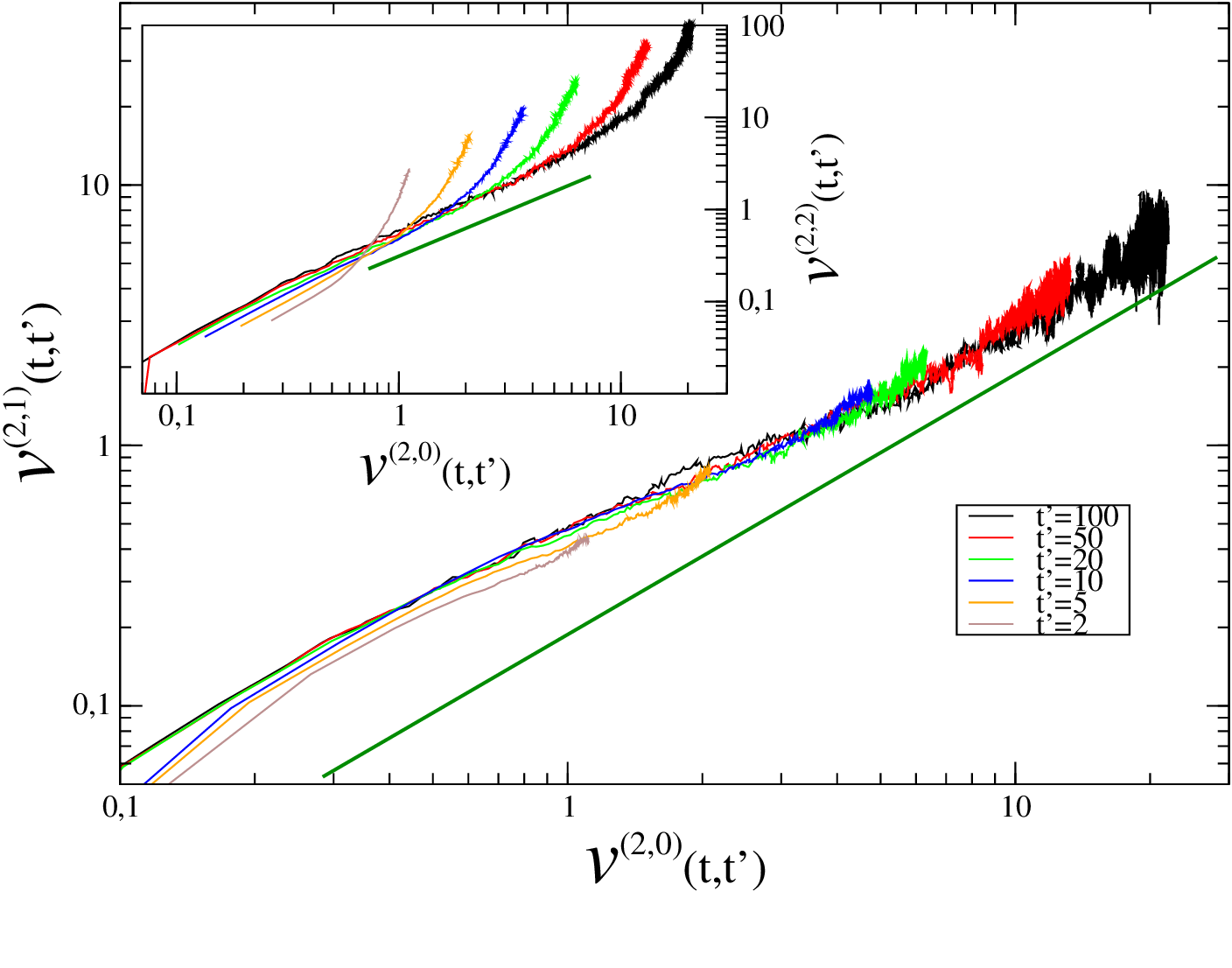}}}
   \vspace{1cm}
   \caption{(Color online.) Upper panel: 
Clock model with $p=6$ quenched from $T=\infty$ to $T=0.76$. 
The re-scaled moment $t'^{-a}{\cal V}_{k=0}^{(2,0)}(t,t')$, with $a=0.76$, 
is plotted against $t/t'$ in the main part of the picture, while the same plot
for ${\cal V}_{k=0}^{(2,1)}$ (with a minus sign in order to have a positive quantity)  
and ${\cal V}_{k=0}^{(2,1)}$ is presented in the
upper and lower insets, respectively. Different choices of $t'$
correspond to different curves, see key in the figure. 
Lower panel: Same data as in the upper panel
but plotted in the parametric form $-{\cal V}_{k=0}^{(2,1)}(t,t')$ against 
${\cal V}_{k=0}^{(2,0)}(t,t')$ (main part) and ${\cal V}_{k=0}^{(2,2)}(t,t')$
against ${\cal V}_{k=0}^{(2,0)}(t,t')$ (inset). The straight green bold line is the linear
behavior (i.e. scaling functions are proportional).}
\label{fig-p6-quenched}
\end{figure}

The second moments ${\cal V}_{k=0}^{(2,0)}(t,t')$, ${\cal V}_{k=0}^{(2,1)}(t,t')$
and ${\cal V}_{k=0}^{(2,2)}(t,t')$
are plotted in Fig.~\ref{fig-p6-quenched}.
In this figure we used the same scaling procedure for the
heated system, with the same exponent $a$ since, being an equilibrium quantity
it should not depend upon the non-equilibrium protocol. 
As it is seen, this produces a quality of data collapse
comparable to the case of the heated system.
Strictly speaking, one should expect logarithmic correction to this scaling behavior,
caused by the presence of vortices~\cite{vortici}. However, such corrections are very
tiny in the asymptotic time domain and they cannot be detected from the inspection of
our data.
A major difference with respect to the heated system is, instead, the behavior
of the scaling functions. Indeed, while $f^{(2,2)}$ keeps growing with the same
power law of the heated system (the two quantities - in the heated and quenched case -
are basically indistinguishable
except for small $x$), $f^{(2,0)}$ and  $f^{(2,1)}$ are changed to a much slower,
logarithmic growth. From the parametric plots presented in the lower panel
of Fig. \ref{fig-p6-heated} one argues that $f^{(2,0)}$ and $f^{(2,1)}$ are still
asymptotically proportional. Interesting, in an intermediate time regime also
$f^{(2,2)}$ grows proportionally to the other scaling functions. For larger times,
however, there is a crossover to a faster growth.
This interesting feature shows that  ${\cal V}_{k=0}^{(2,2)}(t,t')$
does not feel the presence of vortices, while the other moments do.
Since these quantities have been proposed~\cite{c4,noivarianze,noichi2,Cocuyo} as efficient tools to detect 
dynamical heterogeneities and to quantify cooperative lengths, our results 
suggest that different lengths are encoded in the different ${\cal V}$'s.
A possible explanation could be  that 
${\cal V}_{k=0}^{(2,0)}$ and ${\cal V}_{k=0}^{(2,1)}$
detect the distance between vortices while ${\cal V}_{k=0}^{(2,2)}$ is only
determined by the typical length of the smooth spin rotations (spin waves), but
this subject should be further investigated.
Notice also that the peculiar scaling of the ${\cal V}_{k=0}$'s found
in the case of the quench has non-trivial consequences on the limiting behavior
of the generalized FD ratios of Eq.~(\ref{xmom-def}).
Indeed, since the ordinary FD ratio $X$ is finite
\cite{noiteff}, the disconnected terms subtracted off in Eq.~(\ref{var})
have the same scaling properties independently on $m$ and $m'$.
Hence the scaling of the ${\cal V}$'s directly inform us on the behavior
of the composite operators ${\cal C} ^{(n,m)}$ (\ref{momC}) and, in turn,
of the ratios $X^{(n;m,m')}$ of Eq.~(\ref{xmom-def}). Since the ${\cal V}$'s
are found to scale with the same exponent $a$ but with a different form of the scaling
function this implies that the limiting value $\lim _{t'\to \infty} X^{(2;m,m')}(t=yt',t')$
(with $y$ fixed) is finite. The same is true also when the limit 
$\lim _{t\to \infty} X^{(2;m,m')}(t,t')$ is taken (with $t'$ sufficiently large) but only
for $m\neq 2$ and $m'\neq 2$. On the other hand, the same quantity is not
finite for $m=2$ or $m'=2$. This behavior is radically different from all the other 
cases studied insofar.

\vspace{1.5cm}

\section{Conclusions}
\label{sec:conclusions}

In this paper we undertook the study of 
the out of equilibrium  dynamics of some unfrustrated models 
with a finite FD ratio from the novel
perspective of dynamic fluctuations.
After defining in a proper way the fluctuating quantities the average of which 
yield the usual two-time correlation and linear response function,
we evaluated the properties of their probability distribution
and the scaling of the composite operators [defined in Eqs.~(\ref{momder}) and (\ref{momC})]
made of products of $n$ of these fluctuating quantities. 
We showed that, in the model studied analytically, 
such composite fields in the asymptotic time domain scale 
in the same way when the total number $n$ of fluctuating parts involved is the same, 
irrespectively of how many factors are of the correlation or 
the linear response type. 
Therefore ratios between composite operators with the same $n$
converge to a finite value in the large time limit for any value of $n$.
Since such composite operators are strictly related to
higher order correlation and response functions, these ratios can be
regarded as a generalization of the usual FD ratio 
above linear order. 
In the restricted context of the simple models considered analytically in this Article,
their finite asymptotic value 
might speak about the significance of the notion
of an effective temperature associated to the FD ratio. 
Indeed, for such a concept to be physically meaningful, one would ask
such effective temperature to remain finite at any order.
Related to that, the analytical results of this Article support the idea that fluctuations in 
aging systems are intimately related to the behavior of the FD ratio~\cite{ChamonCugliandolo,Cocuyo}.
The mechanism whereby a finite value of the generalized FD ratios is attained in the simple unfrustrated 
models considered analytically here might also be useful to understand the behavior of 
fluctuations in more realistic systems.

Besides, the analysis of this Article is also related to
the problem of the detection of characteristic lengths form higher order
correlations and response functions \cite{Cocuyo,noichi2,lengths}. 
One major problem in this context is which lengths are these higher order 
quantities sensitive to, and how. Interestingly enough, in all the models
considered in this Article where a unique growing length is present
the composite operators have the same asymptotic time-scaling 
(exponents and scaling functions). This indicates that the growing length 
enters different composite operators in the same scaling way.
The different scaling functions found in the quenched clock model,
where two different lengths are present, seems to indicate that
different composite operators are sensitive to different lengths. 

Some previous studies of the second-order momenta in different model 
systems had been performed and we wish now to confront our findings to these. 
In so doing, we will refine the picture of the dynamic scaling of fluctuations in i) coarsening 
systems quenched below their critical temperature; ii) disordered 
spin models such as the Edwards-Anderson spin-glass; iii) critical 
systems as the ones we studied here.

The results for the critical cases considered here (except possibly from the quenched clock model) 
are clearly different from what has been found 
in sub-critical quenches of simple coarsening systems~\cite{noivarianze} where 
the moments associated to the correlation scale differently from the ones 
where the fluctuations associated to the linear response enter. The analysis of the 
second-order momenta for a ferromagnet quenched to its critical point 
computed numerically in \cite{noivarianze} and analytically
in the spherical model in~\cite{annibsollich} unveil a behavior analogous
to the one found in this paper and the scaling found by these authors 
conforms to the form given in Eq.~(\ref{eq:general-scaling}).
Monte Carlo simulations of the sub-critical dynamics of the 
$3d$ Edwards-Anderson  spin-glass~\cite{noispinglass}
suggest that the second moments in these glassy systems scale in the 
same way, in agreement with the conclusions arrived at in~\cite{Chkecacu,Cachcuke,Cachcuigke,Jaubert}
by analyzing the joint probability distribution of Sec.~\ref{sec:joint}.
This claim has to be taken with the usual proviso that numerical simulations 
of glassy systems are hard to interpret beyond any doubt.

The analysis in this paper could be applied to other systems with glassy dynamics;
thus helping to complete a general comprehension of fluctuations in problem with  
slow dynamics. Obvious candidates are kinetically constrained models~\cite{kinetically-constrained}, 
for which an analysis of out of equilibrium fluctuations along these lines was initiated in~\cite{Chchcurese};
the one dimensional Glauber Ising model~\cite{Garrahan}, or random manifold problems, studied 
from an averaged perspective in~\cite{Schehr} among many other papers.

\appendix

\section{Calculation of ${\cal C}_\lambda^{(n,0)}$} 
\label{app1}

Using the exponential form of the cosine we have
\be
{\cal C}_\lambda^{(n,0)}=2^{-n}\langle \prod_{i=1}^{n}\left (e^{i\delta _{i,\lambda}}+
e^{-i\delta_{i,\lambda}}\right )\rangle =
2^{-n}\sum _{\{s_i = \pm 1\}}\langle e^{i\sum _{i=1,n}s_i \delta _{i,\lambda}} \rangle
\; ,
\ee
where $s_i=\pm 1$ are {\it sign} variables.
Using the property 
$\langle \exp(\pm ia)\rangle=\exp(-\langle a^2\rangle/2)$, holding
for any linear function $a(\xi)$ of $\xi$, since the $\delta _{i,\lambda}$ are
themselves linear, one arrives at 
\be
{\cal C}_\lambda^{(n,0)}=2^{-n}\sum _{\{s_i= \pm 1\}}
e^{-\frac{1}{2}\langle \left (\sum _{i=1,n}s_i \delta _{i,\lambda}\right )^2 \rangle}
=2^{-n}\sum _{\{s_i = \pm 1\}} e^{-\frac{1}{2}\langle {\cal S}_{\lambda}^2 \rangle }
\; ,
\ee
where the last equality defines the function $ {\cal S}_{\lambda}$.
We are now able to compute the $n$-times derivatives
involved in Eq.~(\ref{momder}). 
By taking them one at a time it is easy to check that
$\partial _{\lambda _1} ... \partial _{\lambda _n}
e^{-\frac{1}{2}\langle {\cal S}_{\lambda} ^2 \rangle }=\left (\prod_{i=1}^{n}s_i\right )
P^{(n)}({\cal S} _{\lambda},\{ \partial \alpha _{\lambda_i}\} )
e^{-\frac{1}{2}\langle {\cal S}_\lambda^2 \rangle }$, where the polynomial $P^{(n)}$
can be obtained by the recursive relation
\be
P^{(r)}({\cal S} _\lambda ,\{\partial \alpha _{\lambda_i}\})=-P^{(r-1)}({\cal S} _\lambda,\{\partial \alpha _{\lambda_i}\})
\langle {\cal S}_{\lambda} \partial \alpha _{\lambda_r}\rangle
+s_r^{-1}\partial _{\lambda_r}P^{(r-1)}({\cal S} _\lambda,\{\partial \alpha _{\lambda_i}\})
\; ,
\label{recursive}
\ee 
starting from $P^{(0)}({\cal S} _\lambda ,\{\partial \alpha _{\lambda_i}\})\equiv 1$.
Hence one has
\be
\partial {\cal C}^{(n,0)}_{\{\lambda \}=0}=2^{-n}\sum _{\{s_i=\pm 1\}}\left (\prod_{i=1}^{n}s_i\right )
P^{(n)}({\cal S},\{\partial \alpha _{\lambda_i}\})
e^{-\frac{1}{2}\langle {\cal S}^2 \rangle }
\; ,
\label{dergen1}
\ee 
where ${\cal S} = {\cal S} _{\{\lambda =0\}}$.
The recursive equation (\ref{recursive})
implies that $P^{(n)}$ is a sum which contains only
products of correlators of the type 
\be
f=\langle {\cal S} \partial \alpha _{\lambda_i}\rangle
\; ,
\ee
and 
\be
\overline f=\langle  \partial \alpha _{\lambda_i} \partial \alpha _{\lambda_j}\rangle
\; .
\label{f2}
\ee
Indeed, the first term in the rhs of the recursive equation (\ref{recursive}) 
is itself a multiplication by a factor of type $f$,
while the second term amounts to the replacement of some terms of type
$f$ with others of type $\overline f$ (because  
$s_j^{-1}\partial _{\lambda_j} \langle {\cal S} \partial \alpha _{\lambda_i}\rangle
=\langle  \partial \alpha _{\lambda_i} \partial \alpha _{\lambda_j}\rangle$). 
It is easy to show (see App.~\ref{app2}) that in a large time limit 
[for $t'\to \infty$, with $y$ and $r$ (or $\zeta$) finite,
or equivalently for $t-t'\to \infty$ with $r$ (or $\zeta$) finite]
the terms of type $\overline f$ are sub-dominant.
Hence, in this time sector one has 
$P^{(n)}\simeq P^{(n)}(\{\langle {\cal S} \partial \alpha _{\lambda _i}\rangle \})$.

\section{Asymptotic analysis} 
\label{app2}

Terms of kind $f$ are sums (over $i$) of contributions of the form
\be
f(x_i,x'_j;t_i,t'_j)=\langle \phi (x_i,t_i)
\partial _{t'_j}\phi (x'_j,t'_j)\rangle=
\partial _{t'_j} C_\phi(\vert x_i-x'_j\vert;t_i,t'_j)
\; , 
\label{fa}
\ee
for $j\leq m$, or     
\be
f(x_i,x'_j;t_i,t'_j)=\frac{1}{2T}\langle \phi (x_i,t_i)
\xi (x'_j,t'_j)\rangle=R_\phi(\vert x_i-x'_j\vert;t_i,t'_j) 
\label{fb}
\ee     
for $j>m$. Next to these there are analogous contributions that are obtained 
from the terms (\ref{fa}) and (\ref{fb}) with the replacement
$(\vec x_i,t_i)\to(\vec x_i',t_i')$.

As it can be seen from the recursive equation (\ref{recursive}), 
at the $r$-th step, terms of type $\overline f$ are generated from
the quantities of type $f$ already present at step $r-1$ by replacing ${\cal S} $
with a $\partial \alpha _{\lambda _r}$.
Then, in a generic term $\overline f$ it is always $i\neq j$. 
Using Eq.~(\ref{propR}) for the quantities of type $\overline f$ with $i,j>m$, 
one has  $\overline f(x'_i,x'_j;t'_i,t'_j)=(2T)^{-2}\langle \xi(x'_i,t'_i) \xi(x'_j,t'_j)\rangle$,
which, restricting to the case, $(x'_i,t'_i)\neq(x'_j,t'_j)$ $\forall ij$, vanishes identically. 
On the other hand, for $i,j\le m$, using Eq.~(\ref{propdC})
one has 
\be
\overline f(x'_i,x'_j;t'_i,t'_j)=\langle \partial _{t'_i}\phi (x'_i,t'_i)
\partial _{t'_j}\phi (x'_j,t'_j)\rangle=
\partial _{t'_i}\partial _{t'_j} C_\phi(r_{ij};t'_i,t'_j)
\; , 
\ee     
and for $i\le m$ and $j>m$ 
\be
\overline f(x'_i,x'_j;t'_i,t'_j)=\frac{1}{2T}\langle \partial _{t'_i}\phi (x'_i,t'_i)
\xi (x'_j,t'_j)\rangle=
\partial _{t'_i} R_\phi(r_{ij};t'_i,t'_j) 
\; .
\ee     
Comparing with the corresponding terms (\ref{fa}) and (\ref{fb}) of type $f$,
there is an extra time derivative in the quantities  $\overline f$.
Using the scaling properties (\ref{eq:bare-dC-scalar}) and (\ref{eq:bare-R01-scalar}) one
concludes that the latter are negligible in the large times domain
[for $t'\to \infty$, with $y$ and $r$ (or $\zeta$) finite,
or equivalently for $t-t'\to \infty$ with $r$ (or $\zeta$) finite].

\section{Scaling of $C_\phi(r;t,t')$} 
\label{app3}

From Eq.~(\ref{eq:gamma-scalar}), for $2\Lambda ^2(t-t')\gg 1$ one can write
\be
C_\phi(r;t,t')=\frac{T}{4\pi}\int ^\zeta _{\zeta\left [\frac{y-1}{y+1}\right]}
dz \ \frac{e^{-z}}{z}
\ee
where $\zeta=r^2/[2(t-t')]$ and $y=t/t'$.
For $\zeta \ll 1$, neglecting the exponential factor one obtains
\be
C_\phi(r;t,t')=\frac{T}{4\pi}\ln \left (\frac{y+1}{y-1}\right),
\ee
For larger values of $\zeta$, instead, letting $y\gg 1$ one finds 
\be
C_\phi(r;t,t')=\frac{T}{2\pi}(y-1)^{-1}e^{-\zeta}.
\ee

Finally, we consider the equal times correlation. Proceeding as before we write
\be
C_\phi(r;t,t)=\frac{T}{4\pi}\int ^{\Lambda^2 r^2} _{\frac{\Lambda^2 r^2}{4\Lambda ^2t+1}}
dz \,\frac{e^{-z}}{z}
\; . 
\ee
For $\Lambda^2 r^2 \ll 1+4\Lambda ^2 t$, again neglecting the exponential, one has
\be
C_\phi (r;t,t)=\frac{T}{4\pi}\ln(1+4\Lambda ^2 t)
\ee
For $1<\Lambda^2 r^2<1+4\Lambda ^2 t$ one can split the integral as
\be
C_\phi(r;t,t)=
\frac{T}{4\pi}
\left\{ 
\int ^{1} _{\frac{\Lambda^2 r^2}{4\Lambda ^2t+1}}
\,\frac{e^{-z}}{z}\,+\,
\int ^{\Lambda^2 r^2} _1
\,\frac{e^{-z}}{z}
\right\}
\; ,
\label{split}
\ee
and hence
\be
C_\phi(r;t,t)\simeq \frac{T}{4\pi}\ln \left (\frac{1+4\Lambda^2 t}{\Lambda^2 r^2}\right )
\; ,
\ee
since the second integral in Eq.~(\ref{split}) is exponentially suppressed.
Finally, for $\Lambda^2 r^2\gg 1+4\Lambda ^2t$, $C_\phi (r;t,t)$ is exponentially small.
Putting everything together one has
\be
C_\phi(r;t,t)\simeq  \frac{T}{4\pi}
\ln \left (\frac{1+4\Lambda^2 t}{\Lambda^2 r^2+1}\right )
f_{cut}\left (\frac{\Lambda^2 r^2}{1+4\Lambda ^2 t}\right )
\; ,
\ee
where $f_{cut}(z)$ is a cut-off function.

\vspace{2cm}

\noindent
{\bf Acknowledgments}
L. F. C. wishes to thank Federico Rom\`a and Daniel Dom\'{\i}nguez for early discussions on this
problem. This work was financially supported by ANR-BLAN-0346 (FAMOUS).

\end{document}